# Astroparticle yield and transport from extragalactic jet terminal shocks

Fabien Casse [a] & Alexandre Marcowith [b]

[a] *FOM - Institute for Plasma Physics "Rijnhuizen", PO Box 1207 NL-3430 BE Nieuwegein, Netherlands*

[b] *C.E.S.R., 9 avenue du colonel Roche, BP 4346, F-31028 Toulouse, France*

**Abstract**

The present paper deals with the yield and transport of high-energy particle within extragalactic jet terminal shocks, also known as hotspots. These astrophysical sources are responsible for strong non-thermal synchrotron emission produced by relativistic electrons accelerated via a Fermi-type mechanism. We investigate in some details the cosmic ray, neutrinos and high-energy photons yield in hotspots of powerful FRII radio-galaxies by scanning all known spatial transport regimes, adiabatic and radiative losses as well as Fermi acceleration process. Since both electrons and cosmic rays are prone to the same type of acceleration, we derive analytical estimates of the maximal cosmic ray energy attainable in both toroidal and poloidal magnetic field dominated shock structures by using observational data on synchrotron emission coming from various hot-spots. One of our main conclusions is that the best hot-spot candidates for high energy astroparticle production is the extended ($L_{HS} \geq 1 kpc$), strongly magnetized ($B > 0.1 mG$) terminal shock displaying synchrotron emission cut-off lying at least in the optical band. We found only one object (3C273A) over the six objects in our sample being capable to produce cosmic rays up to $10^{20}$ eV. We also show that the Bohm regime is unlikely to occur in the whole hot-spot since it would require unrealistically low jet velocities. Secondly, we investigate the astroparticle spectra produced by two characteric hot-spots (Cygnus A and 3C273 A) by applying a multi-scale MHD-kinetic scheme, coupling MHD simulations to kinetic computations using stochastic differential equations. We show that 3C273 A, matching the previous properties, may produce protons up to $10^{20}$ eV in a Kolmogorov type turbulence by both computing electron and cosmic ray acceleration. We also calculate the high-energy neutrino and gamma-ray fluxes on Earth produced through p-$\gamma$ and p-p processes and compare them to the most sensitive astroparticle experiments.

## 1 Introduction

The origin of ultra high-energy cosmic rays (UHECRs) with energies beyond $10^{19}$ eV is still uncertain. A consensus stands for an extragalactic origin based on several arguments: i/ the acceleration and confinement of UHECRs in our galaxy is difficult to obtain due to the weakness of the galactic magnetic field; ii/ the global isotropy of the high-energy events observed by the ground based experiments AGASA & HIRES (even if some multiplets events have been reported by Takeda et al (1999)); iii/ an indication for a non-scale invariant spectrum at energies $\simeq$ 30 Eev (1 Eev = $10^{18}$ eV) (associated with a bump) and iv/ a possible roll off beyond due to the GZK cut-off (the main scientific task of the AUGER experiment).

Few astrophysical sources seem to be suited for accelerating relativistic particles up to such energies. We can cite gamma-ray bursts (Waxman (1995), Gialis & Pelletier (2004) and references therein), shock waves in large scale structures (Norman et al, 1995) and active galactic nuclei. The terminal shocks of Fanaroff-Riley type II radio galaxies jets are among the most extended and powerful shocks in the universe and are known to be efficient accelerators. They exhibit non-thermal signatures in the radio through infra-red or even optical wave bands (see the review by Meisenheimer (2003) and references therein). A growing number of hot-spots are now detected in X-rays (see Brunetti et al (2001)). This emission, even if interpreted via different mechanisms (synchrotron radiation from electrons or protons, Compton Inverse on cosmic or synchrotron photons), requires high-energy particles. Consequently it does not seem incoherent to consider these sources as possible sites for the production of UHECRs with energies up to $E_{max} \simeq 100$ EeV.

Following a previous work by Biermann & Strittmatter (1987), Rachen & Biermann (1993) calculated the maximal energies that protons can reach in hot-spots under the assumption of a Kolmogorov turbulence accounting for radiative (through synchrotron radiation and p-$\gamma$ interaction) and added the effect of escape losses. They constrained the turbulence properties with the help of the synchrotron radiation produced by the electrons accelerated in the same conditions. The model parameters were derived from observations of the synchrotron cut-off as well as the hot-spots linear size and radius (Meisenheimer et al (1989), and Meisenheimer et al (1997)). The authors concluded that diffusive shock theory can explain the main hot-spot features and is able to accelerate cosmic rays up to 100 EeV. However their work suffers from some large uncertainties in the derivation of $E_{max}$ amongst the shock wave obliquity and the turbulence downstream. In this paper, in order to improve upon these uncertainties, we reconsider the work of Rachen & Biermann (1993) and test the UHECRs acceleration versus all *known* isotropic turbulence scalings, namely Kolmogorov, Kraichnan (we provide detailed calculations for these first two scalings), Bohm, anisotropic turbulence (Goldreich & Sridhar ,

*Email addresses:* fcasse@cdf.in2p3.fr ; Now at PCC -Collège de France, 11 place M.Berthelot F-75231 Paris Cedex 05, France,
Alexandre.Marcowith@cesr.fr (Alexandre Marcowith).

1995) and physical conditions (hot-spot extension, magnetic field strength).

The particle distribution resulting from energy dependent spatial random walks coupled with the Fermi processes and radiative losses can only be calculated numerically. Following a previous work (Casse & Marcowith , 2003), we used coupled 3D axisymmetric MHD simulations and stochastic differential equations (SDE) computations to derive the distribution of cosmic rays escaping the hot-spot and distributions of both high-energy neutrinos and gamma-rays produced during $p - \gamma$ and $p - p$ interactions within the source. Due to the large extension of the hot-spot, the conditions required to accurately compute diffusive shock acceleration with the SDE method are difficult to fulfill unless an adaptative mesh refinement (AMR) algorithm is used. We present then the first multi-dimensional AMR-MHD simulations of hot-spots coupled with kinetic scheme calculations suitable for relativistic particles transport.

The article is organized as follows: in section 2 we present the combined MHD-SDE approach and the main physical processes relevant for relativistic particle production in hot-spots; in section 3, we discuss the constraints on particle acceleration considering different kind of turbulent spectra and section 4 is devoted to numerical simulations of different classes of hot-spots, and their high-energy particle yield.

## 2  Theoretical approach: MHD and kinetic theory

We have adopted the same approach as Casse & Marcowith (2003) where both macroscopic (MHD) and microscopic descriptions (kinetic) are considered. We shall first detail the kinetic scheme used in these simulations for both electrons and cosmic ray transport. We shall then present the numerical code describing the temporal evolution of a magnetized extragalactic jet propagating in a dense medium. As a final step, we shall explain the initial configuration of the system as well as the fiducial quantities used to normalize physical quantities.

### 2.1  Kinetic description: SDE

Kinetic transport of non-thermal particles mainly relies on Fokker-Planck type equations. These equations determine temporal evolution of the distribution function, $f$, including physical transport processes acting on various kind of particles such as ions, electrons or neutrinos. When such non-thermal particles are embedded in a thermal fluid propagating with a velocity **u**, the Fokker-Planck equation describing the evolution of the distribution function can be written in a very general way as

$$\frac{\partial f}{\partial t} = -(\mathbf{u} \cdot \nabla)f + \frac{1}{3}(\nabla \cdot \mathbf{u})p\frac{\partial f}{\partial p} + \nabla_i.(D_{ij}\nabla_j f)$$
$$+ \frac{1}{p^2}\frac{\partial}{\partial p}\left(D_{pp}p^2\frac{\partial f}{\partial p} + \sum_k p^3 \frac{f}{t_{loss,k}(p)}\right) \quad (1)$$

where $p$ is the particle momentum and $t_{loss}(p)$ accounts for characteristic timescales of the energy loss process $k$. Diffusion coefficients $D_{ij}$ and $D_{pp}$ stand for spatial and energy diffusion. When a shock occurs within the thermal plasma, the local velocity field exhibits a very negative velocity divergence $\nabla \cdot \mathbf{u}$. The interaction of non-thermal particles with such a shock results in a Fermi-type acceleration process (Jones & Ellison , 1991).

### 2.1.1 Fermi acceleration theory

In the first-order Fermi acceleration theory framework, following Blandford & Ostriker (1978), let us assume the presence of a plane shock characterized by a compression ratio relating upstream and downstream media quantities $r = \rho_d/\rho_u = U_u/U_d$ ($U_d$ and $U_u$ are assumed constant). For the case where no energy losses occur in the system, the steady-state Fokker-Planck equation (1) reads

$$u\frac{\partial f}{\partial x} - \frac{\partial}{\partial x}\left(D\frac{\partial f}{\partial x}\right) = (U_d - U_u)\delta(x)\left\{\frac{\partial f}{\partial \ln p^3}\right\} \quad (2)$$

where $x$ is the spatial coordinate along the direction normal to the shock (located at $x = 0$) and $D = D(x, p)$ the spatial diffusion coefficient along $x$. On every side of the shock, the flux $uf - D\nabla f$ has to be constant so that the general solution for the spatial part of the isotropic distribution function $f(x, p) = f_1(x)f_2(p)$ in the upstream medium is

$$f_1(x)f_2(p) = \frac{A_1(p)}{U_u} + A_2(p)exp\left(-\int_0^x \frac{U_u dx'}{D(x', p)}\right) \quad (3)$$

The two functions $A_1$ and $A_2$ can be determined thanks to the boundaries at the shock $f_1(x = 0)f_2(p) = f_S(p)$ and at the source outer-edge $f_1(x = L_S)f_2(p) = f_{inj}(p)$. For the case where $L_S$ is large enough to insure $\int_0^{L_S} U_u dx'/D(x', p) \gg 1$, the solution becomes trivial so one can get (Blandford & Ostriker, 1978)

$$f_1(x)f_2(p) = f_{inj}(p) + (f_S(p) - f_{inj}(p))exp\left(-\int_0^x \frac{U_u dx'}{D(x, p)}\right). \quad (4)$$

From this expression we now have $\partial f/\partial x$ and since the particle energy flux $-u\partial f/\partial \ln p^3 - D\nabla f$ has to be continuous throughout the shock discontinuity, the particle energetic spectrum measured at the shock front will be a solution of $df_S/d\ln p^3 = (f_S - f_{inj})U_d/(U_u - U_d)$, namely

$$f_S(p) = -qp^q \int_0^p f_{inj}(p') p'^{-q-1} dp' \quad (5)$$

where $q = -3r/(r-1)$. The particle injection distribution function $f_{inj}$ is typically a function having an upper limit such that $f_{inj}(p > p_o) = 0$ where $p_o$ is considered as the particle momentum injection. For momentum $p > p_o$ the above spectrum will thus be a power law whose index is solely controlled by the shock compression ratio $r$. We have to keep in mind that this statement is valid as long as the condition $\int_0^{L_S} U_u dx'/D(x', p) \gg 1$ is fulfilled, namely if the typical particle diffusion length of momentum $p$ is much smaller than the size of the source. This result is important because it ensures that at the energy where $L_S \gg D(p)/U_u$ the shape of the spectrum does not depend on spatial diffusion coefficients. On the other hand, when particles reach energies where $L_S \sim D(p)/U_u$ the above-mentioned result no longer holds true so that the spectrum will no longer be a power law but a curve with a rapidly decreasing slope. This effect is the translation of source particle leaks competing with the Fermi acceleration.

It is noteworthy that energy diffusion can also occurs within astrophysical objects when turbulence is occurring ($D_{pp} \neq 0$): inelastic scattering of particles by Alfvén waves leading to an increase of the variance of $f_2(p)$ proportional to the square root of time. This indirect acceleration is often called second order Fermi acceleration.

### 2.1.2 Numerical approach

Our kinetic approach is based on the use of stochastic differential equations (SDE) whose structure is very close to Fokker-Planck equations (Krülls & Archterberg, 1994). As already pointed out by Casse & Marcowith (2003), equation (1) can be written in an axisymmetric framework $(R, \theta, Z)$ as

$$\begin{aligned}
\frac{\partial F}{\partial t} = &-\frac{\partial}{\partial R}\left(F\left\{u_R + \frac{\partial D_{RR}}{\partial R} + \frac{D_{RR}}{R}\right\}\right) \\
&- \frac{\partial}{\partial Z}\left(F\left\{u_Z + \frac{\partial D_{ZZ}}{\partial Z}\right\}\right) \\
&- \frac{\partial}{\partial p}\left(F\left\{-\frac{p}{3}\nabla \cdot \mathbf{u} + \frac{1}{p^2}\frac{\partial p^2 D_{pp}}{\partial p} - \sum_k \frac{p}{t_{loss,k}(p)}\right\}\right) \\
&+ \frac{\partial^2}{\partial R^2}(FD_{RR}) + \frac{\partial^2}{\partial Z^2}(FD_{ZZ}) + \frac{\partial^2}{\partial p^2}(FD_{pp}) \quad (6)
\end{aligned}$$

where $F = Rp^2 f$. This equation is strictly equivalent to a set of SDE reading

$$\frac{dR}{dt} = u_R + \frac{\partial D_{RR}}{\partial R} + \frac{D_{RR}}{R} + \frac{dW_R}{dt}\sqrt{2D_{RR}}$$
$$\frac{dZ}{dt} = u_Z + \frac{\partial D_{ZZ}}{\partial Z} + \frac{dW_Z}{dt}\sqrt{2D_{ZZ}} \quad (7)$$
$$\frac{dp}{dt} = -\frac{p}{3}\nabla \cdot \mathbf{u} + \frac{1}{p^2}\frac{\partial p^2 D_{pp}}{\partial p} - \sum_k \frac{p}{t_{loss,k}(p)} + \frac{dW_P}{dt}\sqrt{2D_{pp}}$$

where the $W_j$ are stochastic variables accounting for Wiener processes (see Krülls & Archterberg (1994) for more details). Depending on the nature of the particles, the energy loss phenomena are different. While electrons are prone to synchrotron losses in magnetized plasmas, cosmic rays lose energy via synchrotron radiation, collisions with thermal protons (pp) and photo-disintegration (p$\gamma$) through collisions with ambient photons. We have implemented these energy loss phenomena following Begelman et al (1990).

- *Synchrotron losses*

The typical synchrotron time-scale for electrons is (Rybicki & Lightman, 1979)

$$t_{syn,e}^{-1} = p\frac{4c\sigma_T B^2}{\mu_o m_e c^2} \quad (8)$$

where $c$ is the light velocity, $\sigma_T$ the Thomson cross-section, $m_e$ the electron mass, $B$ the magnetic field and $\mu_o$ the magnetic permittivity of a vacuum. Protons are also prone to this kind of mechanism but with a much lower efficiency since

$$t_{syn,p}^{-1} = \left(\frac{m_e}{m_p}\right)^3 t_{syn,e}^{-1} \quad (9)$$

Inverse Compton (IC) losses can easily be added in the SDE schemes (Eq. (6)) (at least in the Thomson regime). The Inverse Compton (IC) loss time-scale may be expressed in terms of an equivalent magnetic field $B_{eq}$ whose energy density $B_{eq}^2/2\mu_o$ is equivalent to the energy density $U_{ph}$ of the soft photon field involved in IC process. In the present application we can neglect the Compton losses since magnetic energy density is much larger than non-thermal radiation energy density in hotspots. The use of SDE is an interesting method for the computation of synthetic radiative maps at different frequencies (see van der Swaluw & Achterberg (2004) for the supernova remnants case). The spatial extension of one map at a given frequency would then depend on the magnetic fields intensity, the particle energy and then on the dominant radiative process (synchrotron or Inverse Compton). The SDE method, like any other method reconstructing the distribution function of radiating particles, could be employed to discriminate among the physical processes in energetic sources. For instance it could help in a complementary way to spectral studies, to investigate the synchrotron or Inverse Compton dominance of X-ray radiation as observed by Chandra in knots and hot-spots of jets.

• *Inelastic collisions with thermal protons*
The characteristic time-scale of energy loss by collision with thermal (non-relativistic) protons can be written as $t_{pp}^{-1} = n_p \sigma_{pp} K_{pp}$ where $n_p$ is the thermal plasma density, $\sigma_{pp}$ is the collision cross-section and $K_{pp}$ is the inelasticity. The cross-section can be considered as a constant value of $4 \times 10^{-26} cm^2$ and $K_{pp} \simeq 0.5$ (Begelman et al, 1990).

• *Pion photo-production*
When ultra-high-energy cosmic rays encounter photons with energy $\epsilon_\gamma$, they loose their energy through the $\Delta$-resonance provided that cosmic rays have energy beyond the threshold $\epsilon_{p\gamma}$

$$\epsilon_{p\gamma} = 6.6 \times 10^{16} eV \left( \frac{\epsilon_\gamma}{1\ eV} \right)^{-1} \tag{10}$$

$$p + \gamma \xrightarrow{2/3} p + \pi^o \to p + \gamma + \gamma$$
$$\xrightarrow{1/3} n + \pi^+ \to ... \to p + e^+ + e^- + \nu_e + \bar{\nu}_e + \nu_\mu + \bar{\nu}_\mu \ .$$

Since it is observationally proved that relativistic hot-spots electrons are able to produce photons with energies up to optical and in some case X-rays, we can already sense that neutrino and $\gamma$-ray production will be important probes of cosmic ray acceleration. The computation of the characteristic time-scale for photo-disintegration (also called photo-meson production) is not as straightforward as the previous ones because it involves both the proton Lorentz factor $\gamma_p$ and the photon spectrum quantities as spectrum cut-off frequency $\nu_{max}$. As the theory of photo-meson production is not yet able to provide good predictions for the cross-section $\sigma_{p\gamma}$, we have to use experimental data recorded for collisions between protons and very high-energy $\gamma$ photons. These nuclear reactions are similar to the ones considered here if we look at this collision in the proton rest frame. In order to perform the calculation, let us define a measure of photon energy in the rest frame of the proton $x = 2\gamma_p \epsilon_\gamma / m_e c^2$. Using this quantity, we can see that the photo-disintegration energy threshold is $x_{th} \simeq 284$. According to Begelman et al (1990), the time-scale for photo-meson production is

$$t_{p\gamma}^{-1} = \frac{2\pi c}{\gamma_p^2} \int_{2\gamma_p x_{th}}^{2\gamma_p x_{max}} \sigma_{p\gamma}(x^*) K_{p\gamma}(x^*) x^* dx^* \int_{x^*}^{x_{max}} n_x dx \tag{11}$$

where $n_x$ is the photon occupation number defined such as $\int n_x x^2 dx d\Omega$ gives the local density of photons. In order to compute this time-scale, we need to have the photon distribution. This distribution will be provided by first computing the electron acceleration and the related synchrotron spectrum. For the case of power-law photon spectrum, Aharonian (2002) provides a useful estimate of $t_{p\gamma} \sim 10^9 c_\alpha S_o^{-1} (\epsilon_p/10^{19}eV)^{-\alpha}\theta_{obs} yr$, where $c_\alpha$ is a constant or order unity depending on the power law index $\alpha$, $S_o$ is the observed flux in Jansky and $\theta$ is the angular size of the object in arcseconds. We have performed calculations with both expressions and found that they are similar except at very large energies beyond $10^{19} eV$ where the estimate of Aharonian (2002) is over-predicting the process efficiency.

*2.1.3 Secondary particles spectra*

The collisions of cosmic rays with protons and photons of the astrophysical source produce secondary particles as electron-positron pairs, neutrinos and $\gamma$-rays provided that cosmic rays have energy beyond the reaction energy threshold. For the inelastic collision with photons, the energy threshold arises from the requirement that the total energy in the center of mass frame must be high enough to create pions (see Eq.(10)). The neutrinos and gamma-ray fluxes are given by

$$\frac{dF_\gamma}{dt} = \frac{4c}{3} \int F_{ph} \int_{\epsilon_{min}}^{\infty} F_{CR}(\gamma_p)\sigma_{p\gamma}\delta(\epsilon_\gamma - \bar{\epsilon}_\gamma)d\epsilon_{ph}d\gamma_p$$

$$\frac{dF_\nu}{dt} = \frac{4c}{3} \int F_{ph} \int_{\epsilon_{min}}^{\infty} F_{CR}(\gamma_p)\sigma_{p\gamma}\delta(\epsilon_\nu - \bar{\epsilon}_\nu)d\epsilon_{ph}d\gamma_p \quad (12)$$

where $\bar{\epsilon}_{\gamma/\nu}$ is the $\gamma$-ray (neutrino) energy averaged over the angle $\varphi$ and the functions $F$ are defined as $\epsilon^2 f$ (as for instance $F_{CR} = \gamma_p^2 f(\gamma_p)$).
The factors 4/3 in Eq. (12) have different origins. The interaction p-$\gamma$ produces first a neutral pion at a rate of 2/3 which further decays in 2 photons and then produces a charged pion at a rate 1/3 which produces 4 neutrinos flavors. Note also that the cross section $\sigma_{p\gamma}$ is a very peaked function centered near the reaction threshold such that a good approximation is $\sigma_{p\gamma} \sim 5 \times 10^{-28}$ cm$^2$ H($\sqrt{s} - m_\Delta c^2 + \Gamma_\Delta/2$)H($m_\Delta c^2 + \Gamma_\Delta/2 - \sqrt{s}$) where the terms contained in the Heaviside function H are $\sqrt{s}$ the system energy measured in the center of mass frame, $m_\Delta c^2 = 1.232 GeV$, and the width of the $\Delta$-resonance $\Gamma_\Delta = 0.11 GeV$.
The $\gamma$-rays produced by the disintegration of neutral pions have energies depending on the cosmic ray energy $\epsilon = \gamma_p m_p c^2$, the photon energy $\epsilon_{ph}$ as well as the angle between the cosmic ray and the photon $\varphi$. The averaged gamma-ray energy is roughly 10 % of the initial proton energy (half of the pion energy and each pion is produced with a mean energy of the order of 1/5 of the initial proton energy). The p-$\gamma$ gamma-rays are in the range $\sim 10^{16} - 10^{19}$ eV (deduced from the energy threshold given in table 3.2), out of the energy domain of Tcherenkov telescopes. The neutrinos flux produced by charged pion decay is dominated by the muon neutrinos. Following previous reasoning, the averaged energy per neutrino is 5% of the initial proton energy (Stecker, 1979), that is in the range $\sim 10^{15} - 10^{18}$ eV.
Cosmic rays also interact with matter and produce charged and neutral pions and in turn neutrinos and gamma-rays. The neutrino typical energy produced by a proton of Lorentz factor $\gamma_p$ is $\sim 30 \sqrt{\gamma_p}$ MeV $\leq 10^{13}$ eV in our case. These TeV neutrinos are unobservable by the most sensitive experiments (ANTARES and AMANDA) and will not be investigated further. However, the gamma-rays generated by the neutral pion have energies that fall in the energy window of Tcherenkov telescopes and future high-energy gamma-ray missions like AGILE or GLAST. For $E_p^{-2}$ energy density spectra of protons with a non-relativistic minimum energy, the gamma-ray spectrum produced by pion decay peaks at an energy $\simeq m_{\pi^0}/2 \sim$ 70MeV and extends up to $\sim 1/12$ $E_{pmax}$ (1/2 from the neutral pion decay, 1/6 from the pion production in the GeV-TeV domain).

## 2.2 MHD description: AMRVAC code

Among kinetic equations (6), macroscopic quantities need to be obtained in order to perform the kinetic calculation of non-thermal particles. The dynamics of the thermal flow can be obtained from the magnetohydrodynamic theory which is able to describe the temporal evolution of the magnetized fluid. The MHD equations are a combination of fluid and Maxwell's electromagnetic equations. The set of relations expresses the conservation of mass, momentum and energy as well as the magnetic field induction, namely

$$
\begin{aligned}
&\frac{\partial \rho}{\partial t} + \nabla \cdot (\rho \mathbf{u}) = 0 \, , \\
&\frac{\partial (\rho \mathbf{u})}{\partial t} + \nabla \cdot [\rho \mathbf{u}\mathbf{u} + p_{tot} I - \mathbf{B}\mathbf{B}] = 0 \, , \\
&\frac{\partial e}{\partial t} + \nabla \cdot (e\mathbf{u}) + \nabla \cdot (p_{tot}\mathbf{u}) - \nabla \cdot (\mathbf{u} \cdot \mathbf{B}\mathbf{B}) = 0 \, , \\
&\frac{\partial \mathbf{B}}{\partial t} + \nabla \cdot (\mathbf{u}\mathbf{B} - \mathbf{B}\mathbf{u}) = 0
\end{aligned}
\tag{13}
$$

where $\rho$ is the plasma density, $e$ the total internal energy and $P_{tot} = P_{th} + B^2/2$ is the total pressure composed of the thermal and magnetic pressure.
The time integration of the previous equations cannot be done analytically except maybe in some very simple one dimensional cases. Since we intend to use 3D axisymmetric MHD snapshots of the global structure, we used the recent grid-adaptive Versatile Advection Code AMRVAC (see http://www.phys.uu.nl/~toth (Keppens el al., 2003)). This code uses an automated Adaptive Mesh Refinement strategy, where a base grid is refined by adding finer level grids where a higher reso-lution is needed. Finer level grids are adjusted, inserted or removed by periodically checking if the grid structure should be altered in response to the flow dynamics. This procedure allows us to follow shock-dominated or coexisting global and local plasma dynamics accurately in a much more efficient way than with a global refinement of a static grid. We used the robust two-step Total Variation Diminishing Lax-Friedrichs method on all levels. To handle the solenoidal constraint on the magnetic field $\nabla \cdot \mathbf{B} = 0$, our grid-adaptive simulations used a diffusive source term treatment which damps the errors at their maximal rate in accord with the prevailing Courant-Friedrichs-Lewy condition. This was shown to be effective for multi-D AMR MHD simulations (Keppens el al., 2003). The axisymmetric MHD simulations employed six refinement levels reaching an effective resolution of 3200. Refinement was triggered by relative density errors exceeding 1% in a Richardson-type comparison between coarsened and integrated versus integration and coarsened solutions.

### 2.3 Macroscopic hot-spots model

The dynamics of axisymmetric hydrodynamical jet cocoons has been extensively studied over the last decade through numerical simulations intending to describe the interaction of a supersonic jet with a static ambient medium (Massaglia et al, 1996; Komissarov & Falle, 1998). The outcome of these simulations always gives birth to cocoons. These cocoons surround the jet propagating in a self-similar way whether the jet is relativistic or not (Komissarov & Falle, 1998). The morphology of the expanding cocoons depends on both jet and ambient medium properties, namely the sonic Mach number of the jet and the density contrast between the jet and the exterior medium.

The addition of a toroidal magnetic field can completely change this picture. In several works (Lind et al, 1989; Komissarov, 1999), it has been shown that in the case of the occurrence of a strong magnetic field, the head of the jet no longer creates a cocoon but a nose-cone which does not extend into the surrounding medium. This feature exists either for non-relativistic (Lind et al, 1989) or relativistic jets (Komissarov, 1999). It is noteworthy that this jet head configuration is not commonly observed in extragalactic hot-spots so that it is believed that the FRII jets are only carrying a weak magnetic field, namely they are believed to be super-fastmagnetosonic. The aforementioned works only dealt with purely toroidal magnetic field and other authors have found that the presence of a poloidal magnetic field can disrupt the jet confinement when no toroidal field is present because of the creation of an expanding wave (Kössl et al, 1990). However in this study the initial magnetic field configuration was not assuming any jet in the computational domain but only its influence through a prescribed toroidal current that may be conflicting with the propagation of the jet itself.

In order to avoid any interfering initial conditions, we have designed initial conditions where the head of a cylindrical jet is already present in the computational

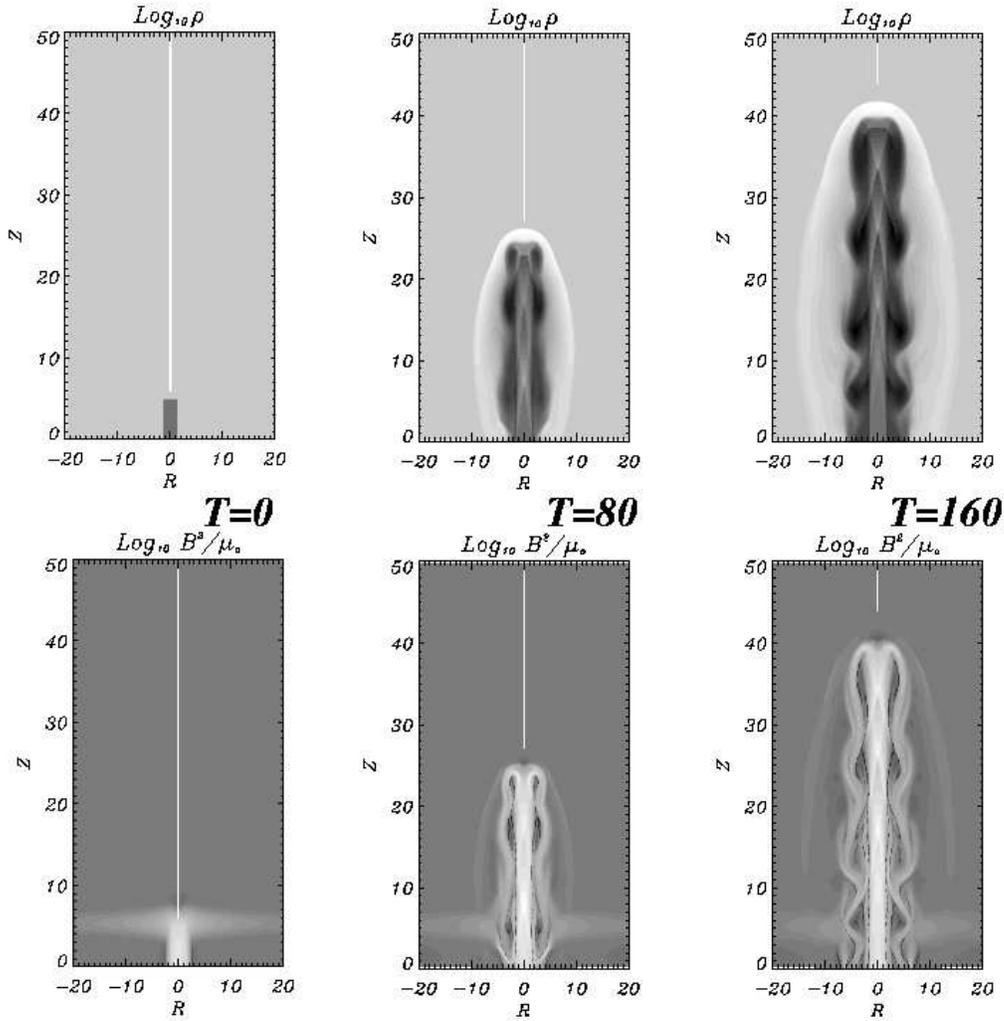

Fig. 1. Snapshots of the logarithmic density (upper panels) and magnetic energy (lower panels) of a magnetized super-fastmagnetosonic jet propagating in a very weakly magnetized dense medium ($\rho_{ext} = 10\rho_{jet}$) at three different stage of the simulation (time **T** is expressed in $R_{jet}/U_{jet}$ units). The formation of a cocoon arises from the jet head propagation where jet material is expelled from the strong shock located at the jet front. This extended shock is believed to host relativistic electrons that are responsible for the strong radio-optical emission seen from these hot-spots. In the present figure the initial magnetic configuration is strictly poloidal, but strictly toroidal simulations lead to very similar structures as long as the jet remains highly super-fastmagnetosonic.

domain and contains a helicoidal magnetic field. The mass density prescription is

$$\rho(R,Z) = \rho_{ext} + \frac{\rho_{jet} - \rho_{ext}}{cosh((R/R_{jet})^{100})cosh((Z/Z_{jet})^{100})} \qquad (14)$$

where $\rho_{jet}$ and $\rho_{ext}$ are respectively the initial jet density and external medium density. The initial extension of the jet is given by $R_{jet}$ and $Z_{jet}$. We assume a constant axial velocity within the jet and nil for the exterior medium (radial velocity is set to zero everywhere). The magnetic field configuration is such that the magnetic energy is contained inside the jet and respect the solenoid nature of a magnetic field. We then display the magnetic field such as

$$B_R(R,Z) = B_o R_{jet}^2 \frac{2(Z/Z_{jet})^3 tanh((Z/Z_{jet})^4) tanh((R/R_{jet})^2)}{R Z_{jet} cosh((Z/Z_{jet})^4)}$$
$$B_Z(R,Z) = \frac{B_o}{cosh((Z/Z_{jet})^4) cosh^2((R/R_{jet})^2)} \quad (15)$$
$$B_\theta(R,Z) = B_1 \frac{(R/R_{jet})}{cosh((Z/Z_{jet})^4) cosh^2((R/R_{jet})^2)}$$

where $B_o$ is the poloidal jet magnetic field strength measured at the axis and $B_1$ controls the field helicity. The value of $B_o$ and $B_1$ will also control the geometry of the terminal shock. Indeed, setting $B_1 = 0, B_o = 1$ will naturally lead to a parallel shock whereas $B_1 = 1, B_o = 0$ will lead to a perpendicular shock. The thermal pressure and toroidal velocity are computed such as the jet is at equilibrium in the radial direction, namely $\partial(P + (B_Z^2 + B_\theta^2)/2)/\partial R = 0$ and $v_\theta = B_\theta/\sqrt{\mu_o \rho}$.

The boundary conditions are open boundaries for $Z = Z_{max}$, $R = R_{max}$ and also at the base of the jet when $R > R_{jet}$. Boundary conditions are frozen to the initial ones at the base of the jet where $R < R_{jet}$. The jet axis is treated as usual with a combination of symmetrical and antisymmetric conditions.

We have performed two kinds of simulation, one being related with jets carrying a purely poloidal magnetic field, the other one being related to a purely toroidal magnetic topology. These two cases represent the two extreme magnetic field configurations that one can expect from astrophysical jets, the real one likely being a combination of both. In both simulations, we have only considered superfastmagnetosonic jets ($M_{S,jet} = 10 = M_{Alfven,jet}$). Indeed, according to typical values of jet magnetic field and density (e.g. Ferrari (1998)), the Alfvén speed in the jet is about $V_A = B/\sqrt{\mu_o \rho} \sim c/100$ and is much smaller than the bulk jet velocity which is believed to be a significant fraction of the speed of light $c(\beta_{jet} \geq 0.1)$. The morphology of the flow is very close to results presented by Lind et al (1989) for weakly magnetized jets and for non-magnetized jets by Massaglia et al (1996). The jet propagates through the denser medium creating a cocoon cavity that isolates the jet from the external medium. The cocoon is created by jet material expelled from the front shock located at the head of the jet. In Fig. 1, we display the logarithmic contours of density at three different stages of the jet propagation. The cocoon clearly appears in the surrounding of the jet, its structure being characterized by a density lower than the jet density as well as an intermediate magnetic field. The propagation of the jet is shown in Fig. 2 where the location of the jet head is plotted as a function of time. This propagation reaches a ballistic motion where the jet head velocity is constant and equal to $0.17 U_{jet}$. This value is lower than the analytical estimates done by Norman et al (1982) where equating the ram pressure of each side of the jet head and assuming no widening of the discontinuity leads to the head velocity $U_h = U_{jet}/(1 + \sqrt{\rho_{ext}/\rho_{jet}}) = 0.24 U_{jet}$. As already pointed out by Lind et al (1989) and Massaglia et al (1996), this value is lower than the previous relationship but is consistent with a jet head widening leading to an increase of the pressure force on one side thanks to the increased shock surface.

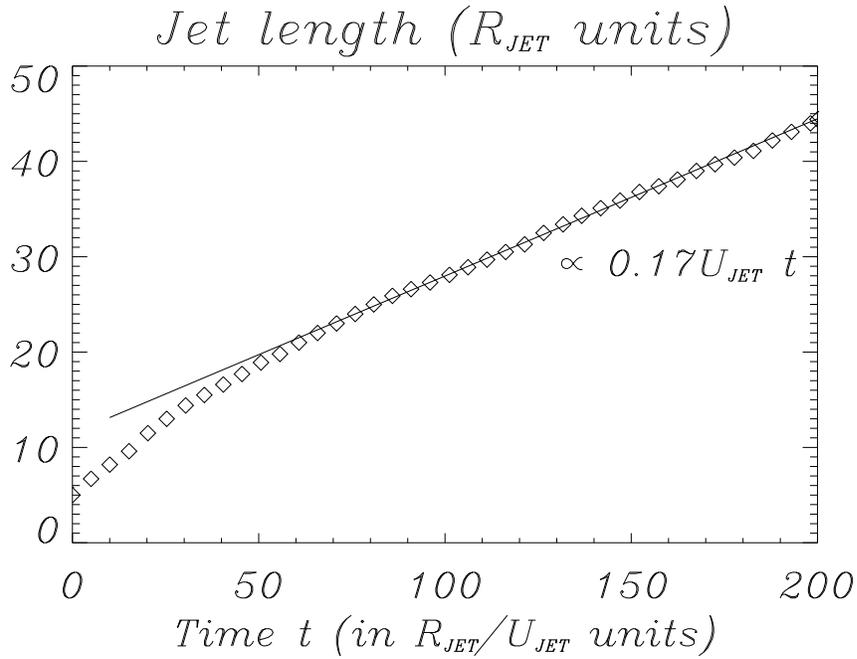

Fig. 2. Jet extension (in $R_{jet}$ units) as a function of time for the simulation presented in Fig. 1. Once arrangements are done between the jet and the external medium, the structure reaches a ballistic motion where the jet head velocity is almost constant and smaller than the inner jet velocity ($U_h \sim 0.17 U_{jet}$).

## 3 Astroparticle yield in hot-spots: Analytical estimates

In this section, we intend to address the issue of high-energy particle production from extragalactic jet terminal shocks. The particles we consider are electrons, synchrotron photons, cosmic rays as well as neutrinos produced by interaction of cosmic rays with the ambient photons. Since acceleration capacities depend highly on the diffusion properties of the plasma, we have considered most of the diffusion regimes identified so far, namely the Kolmogorov and Kraichnan diffusion (perpendicular and parallel shocks), the Bohm diffusion as well as the Goldreich & Sridhar scaling (Goldreich & Sridhar , 1995).

## 3.1 General definitions

When a shock is occurring within an astrophysical plasma, first order Fermi acceleration is taking place. The energy gain as the particle is crossing the shock front can be written as

$$\left(\frac{dp}{dt}\right)_{sh} = p\frac{U_{jet}^2}{r^2 + r}\frac{r-1}{3D} \tag{16}$$

where $D$ is the diffusion coefficient in the direction perpendicular to the front shock and $r$ is the compression ratio of the shock. Posing $\zeta = 3(r^2 + r)/(r-1)$ ($\zeta = 20$ for a strong shock), we can derive the characteristic time-scale for first order Fermi acceleration

$$t_{FI} = \zeta \frac{D}{U_{jet}^2} . \tag{17}$$

Due to the diffusion motion of charged particles, electrons and protons can escape from the astrophysical source. The main particle leakage occurs in the direction with the largest spatial diffusion coefficient. In a general way, we may write the escape time of particle from an axisymmetric object as

$$t_{esc} = min_k\left(\frac{L_k^2}{4D_k}\right) \tag{18}$$

where $k$ indices stand for the various dimensions of the object. In the present work, hot-spots are considered and since they are believed to have roughly the same extension $L_{HS}$ in every directions, we can then write $t_{esc} = min_k(L_{HS}^2/4D_k)$. The synchrotron time-scale has already been defined in equation (8). It is noteworthy that if an electron distribution function exhibits a cut-off energy $\epsilon_c = \gamma_c m_e c^2$ then the associated synchrotron spectrum will also display a cut-off at frequency $\nu_c = 115 B_{0.1mG}\gamma_c^2$ Hz ($B_{0.1mG} = B/0.1mG$).
Spatial diffusion arising from turbulent magnetic fields is still a matter of debate. No theory so far has been able to predict diffusion coefficient expressions coming from compressible MHD turbulence with respect to turbulence level or particle energy, even if some works using numerical simulation of MHD wave spectra have been performed (see Yan & Lazarian (2002) and references therein). Data provided by various spacecrafts measuring solar wind properties are so far the only way to test the different results obtained from theories and/or numerical experiments on diffusion of particles embedded in an astrophysical magnetized plasma prone to magnetic turbulence (see Ruffolo et al. (2003) and references therein). For the case of weak Kolmogorov or Kraichnan-type turbulence, the quasi-linear theory was able to provide a spatial diffusion coefficient along the mean magnetic field as a function of turbulence properties as well as particle energy. Nevertheless its predictions on transverse diffusion has not met success with experiments. Another approach to transverse turbulence, called chaotic magnetic diffusion, is taking into account the diffusive motion of particles and the diffusion of the magnetic field lines (Jokipii, 1969). This theory developed by Rechester & Rosenbluth (1978) and later confirmed by numerical simulations achieved by Casse et al (2002) exhibits spatial transverse diffusion proportional to spatial diffusion along the mean magnetic field $B_o$ through a proportionality coefficient depending only on the turbulence level $0 \leq \eta_T = <\delta B^2>/(B_o^2 + <\delta B^2>) \leq 1$ where $\delta B$ stands for the turbulent component of the magnetic field. The spatial diffusion coefficients are (Casse et al, 2002)

$$D_\parallel = \frac{c\lambda_{max}}{3\pi\eta_T} \tilde{\rho}^{2-\beta},$$
$$D_\perp = \eta_T^{1.3} \frac{c\lambda_{max}}{3\pi} \tilde{\rho}^{2-\beta}. \qquad (19)$$

where $\tilde{\rho} = 2\pi R_L/\lambda_{max} = 2\pi\epsilon/ZeBc\lambda_{max}$ is the reduced particle rigidity ($Ze$ being the particle electric charge) and $\beta$ is the spectral index of the turbulence spectrum ($\beta = 5/3$ for Kolmogorov theory and $3/2$ for Kraichnan one).

In cosmic ray physics, another type of diffusion is considered: the Bohm diffusion. In the limit of strong turbulence ($\eta_T \to 1$), one can assume that a particle mean free path is reduced to its own Larmor radius $R_L$. The natural scaling of a diffusion coefficient naturally leads to $D_B = R_L c/3$ in every direction. Nevertheless, numerical calculations of Casse et al (2002) showed that this Bohm scaling is not in agreement with a Kolmogorov or Kraichnan-type turbulence except if both $\eta_T, \tilde{\rho} \to 1$.

The turbulence could originate from different sources: large scale stochastic fluid motions cascading towards smaller scales and generating magnetic field fluctuations (Pelletier & Zaninetti, 1984), a MHD cascade produced by the relativistic protons themselves (Biermann & Strittmatter, 1987). In the following, we implicitly assumed the second case as the magnetic field energy density will be taken equal to the relativistic particle one, and the maximum scale of the turbulence will be equal to the hot-spot size.

### 3.2 Parallel shocks

The terminology of "parallel" shocks stands for shocks having a magnetic topology such as the mean magnetic field is parallel to the shock normal direction. We will thus consider here that in definitions (17,18) the diffusion coefficient is $D_\parallel$ since this coefficient is always larger than the transverse one in Kolmogorov or Kraichnan theory. In order to determine the energy cut-off $\epsilon_c$ of electrons accelerated at the terminal shock of the jet, we match the first-order Fermi time-scale to the synchrotron time-scale. One then gets

<u>*Kolmogorov*</u> :
$$\epsilon_c = 9.6 \times 10^4 \left(\frac{\eta_T \beta_{jet}^2}{\zeta}\right)^{3/4} \left(\frac{B}{0.1mG}\right)^{-5/4} \left(\frac{\lambda_{max}}{1kpc}\right)^{-1/2} GeV \quad (20)$$

<u>*Kraichnan*</u> :
$$\epsilon_c = 3.6 \times 10^5 \left(\frac{\eta_T \beta_{jet}^2}{\zeta}\right)^{2/3} \left(\frac{B}{0.1mG}\right)^{-1} \left(\frac{\lambda_{max}}{1kpc}\right)^{-1/3} GeV \quad (21)$$

where $\beta_{jet} = U_{jet}/c$. The largest wavelength $\lambda_{max}$ of turbulence spectrum is basically of the order of the size of the hot-spot. The corresponding synchrotron cut-off is then

<u>*Kolmogorov*</u> :
$$h\nu_c = 16.8 \left(\frac{\eta_T \beta_{jet}^2}{\zeta}\right)^{3/2} \left(\frac{B}{0.1mG}\right)^{-3/2} \left(\frac{\lambda_{max}}{1kpc}\right)^{-1} keV \quad (22)$$

<u>*Kraichnan*</u> :
$$h\nu_c = 0.24 \left(\frac{\eta_T \beta_{jet}^2}{\zeta}\right)^{4/3} \left(\frac{B}{0.1mG}\right)^{-1} \left(\frac{\lambda_{max}}{1kpc}\right)^{-2/3} MeV \quad (23)$$

which corresponds to the highest energy photons produced by relativistic electrons through the synchrotron mechanism (see also Biermann & Strittmatter (1987)). The spectrum cut-off then depends both on local ($\eta_T, \zeta, \lambda_{max}$) and macroscopic hot-spot quantities. Acknowledging the uncertainties in the observables, we can see in table 3.2 that the cut-off frequency lies in the range $\simeq 10^{-2}$ to a few eV. Is it possible with the known hot-spot properties to retrieve such a frequency range ? (see for instance Biermann & Strittmatter (1987), Meisenheimer et al (1996) and Wilson & Yang (2002) and references therein, in the context of the M87 jet). One can expect the terminal shocks in FRII jets to be strong and assume $\zeta \simeq 20$. Relying on the observations, it seems reasonnable to take the jet velocity of the order of $0.2 - 0.5c$ and the hot-spot size (assumed to be the maximum turbulence scale) between $10^{-1}$ and a few kpc. These values lead to synchrotron cut-off ranging between $[1 - 100]$ eV $\times (\eta/B_{0.1})^{3/2}$ (in the case of Kolmogorov turbulence, but the following conclusion is more stringent for the Kraichnan turbulence). The turbulence level and the magnetic fields seem then to be the most important parameters to constrain the synchrotron cut-off in hot spots: a cut-off $\ll 1$ eV requires either a low turbulence level or/and a magnetic field well above the equipartition ($\sim 0.3$ mG), inversely, a cut-off of the order of a few eV is consistent with reasonnably high turbulence levels $\geq 0.2$ and magnetic field close to equipartition. Note the fact that the smaller magnetic field, the higher cut-off energy has already been presented by Brunetti et al (2003). Cosmic rays acceleration is also taking place at the terminal shock but contrary to electrons, the main energy loss mechanism is the particle escape from the hot spot. Setting both first-order Fermi and escape time-scale as equal leads to the maximum energy the cosmic rays can reach, namely

$$\underline{Kolmogorov}:$$
$$\frac{\epsilon_{CR,max}}{Z10^{21}eV} = 1.53 \left(\frac{\eta_T \beta_{jet}}{\zeta^{1/2}}\right)^3 \left(\frac{B}{0.1mG}\right)\left(\frac{\lambda_{max}}{1kpc}\right)\frac{L_{HS}^3}{\lambda_{max}^3} \qquad (24)$$

$$\underline{Kraichnan}:$$
$$\frac{\epsilon_{CR,max}}{Z10^{21}eV} = 0.33 \left(\frac{\eta_T \beta_{jet}}{\zeta^{1/2}}\right)^2 \left(\frac{B}{0.1mG}\right)\left(\frac{\lambda_{max}}{1kpc}\right)\frac{L_{HS}^2}{\lambda_{max}^2} \; . \qquad (25)$$

The cosmic ray cut-off can be express in terms of the synchrotron cut-off using equation (22) so that the previous definition is no longer dependent on the turbulence level but on an observational constraint $\nu_c$:

$$\underline{Kolmogorov}:$$
$$\frac{\epsilon_{CR,max}}{Z10^{21}eV} = 1.53 \left(\frac{h\nu_c}{16.8keV}\right)^2 \left(\frac{\zeta^{3/2}}{\beta_{jet}^3}\right)\left(\frac{B}{0.1mG}\right)^4 \left(\frac{L_{HS}}{1kpc}\right)^3 \qquad (26)$$

$$\underline{Kraichnan}:$$
$$\frac{\epsilon_{CR,max}}{Z10^{21}eV} = 0.33 \left(\frac{h\nu_c}{0.24MeV}\right)^{3/2} \left(\frac{\zeta}{\beta_{jet}^2}\right)\left(\frac{B}{0.1mG}\right)^{5/2} \left(\frac{L_{HS}}{1kpc}\right)^2 \; . \qquad (27)$$

The above relations are independent of any local properties of turbulence such as $\eta_T$ or $\lambda_{max}$. The measure of the magnetic field amplitude is a crucial issue here since its value, combined with the observed synchrotron cut-off, directly gives an estimate of the highest energy reachable for ultra-energetic cosmic rays in a given turbulent regime provided that one has an idea of the jet bulk velocity. It is noteworthy that the previous relations are not only valid for hot-spots but for any magnetized astrophysical object exhibiting synchrotron emission coming from a parallel shock vicinity.

Neutrinos astronomy brings new hopes for astroparticle physics. The observation of neutrinos spectra coming from astrophysical environments is of great interest for cosmic rays physics. Indeed these light particles weakly interact with baryonic matter so that it is believed that "neutrino" pictures of a source will be unaltered during neutrinos travel. In hot-spots, the intense synchrotron emission can alter ultra-high energy cosmic rays through the photo-meson production. Looking back at the $\Delta$-resonance threshold in Eq.(10) and considering the maximal cosmic rays energy Eq.(26), we can predict that the hot-spot will be a source of neutrinos if $\epsilon_{CR,max} > \epsilon_{p\gamma}$. The latter inequality can be translated in terms of observational features as

$$\underline{Kolmogorov}:$$
$$\left(\frac{h\nu_c}{1keV}\right) > \left(\frac{\epsilon_{c,min}}{1keV}\right)$$
$$= 2.3 \times 10^{-2} \frac{\beta_{jet}}{\zeta^{1/2}} \left(\frac{B}{0.1mG}\right)^{-4/3} \left(\frac{L_{HS}}{1kpc}\right)^{-1} \quad (28)$$

$$\underline{Kraichnan}:$$
$$\left(\frac{h\nu_c}{1keV}\right) > \left(\frac{\epsilon_{c,min}}{1keV}\right)$$
$$= 5.6 \times 10^{-2} \left(\frac{\beta_{jet}}{\zeta^{1/2}}\right)^{4/5} \left(\frac{B}{0.1mG}\right)^{-1} \left(\frac{L_{HS}}{1kpc}\right)^{-4/5}. \quad (29)$$

| Hot-spot | $\beta_{jet}[c]$ | $B[0.1mG]$ | $L_{HS}[kpc]$ | $\epsilon_{CR,max}^{\beta=5/3}[eV]$ | $\epsilon_{CR,max}^{\beta=3/2}[eV]$ | $h\nu_c[eV]$ | $\epsilon_{p\gamma,min}[eV]$ |
|---|---|---|---|---|---|---|---|
| 3C273 A | 0.27 | 3.6 | 1.9 | $7.9 \times 10^{19}$ | $1.5 \times 10^{17}$ | 1.74 | $3.8 \times 10^{16}$ |
| 3C405A | 0.24 | 3.5 | 1.5 | $4.1 \times 10^{16}$ | $4.9 \times 10^{14}$ | $3.7 \times 10^{-2}$ | $1.78 \times 10^{18}$ |
| 3C405D | 0.3 | 4.1 | 1.4 | $1.6 \times 10^{16}$ | $2.4 \times 10^{14}$ | $3.3 \times 10^{-2}$ | $2 \times 10^{18}$ |
| 3C20 W | 0.47 | 4.8 | 0.13 | $1.2 \times 10^{15}$ | $6.96 \times 10^{13}$ | 0.48 | $1.37 \times 10^{17}$ |
| 3C123 E | 0.4 | 1.9 | 4 | $4.16 \times 10^{13}$ | $3.6 \times 10^{12}$ | $2.5 \times 10^{-3}$ | $2.64 \times 10^{19}$ |
| 3C111 E | 0.35 | 2.4 | 0.07 | $4.5 \times 10^{12}$ | $1.6 \times 10^{12}$ | 0.19 | $3.5 \times 10^{17}$ |

Table 1
Maximum cosmic ray energy attainable within hot-spots for different diffusion regimes (Kolmogorov and Kraichnan) using observed properties given by Meisenheimer et al (1997). In this hot-spot sample, 3C273 A is able to achieve ultra-high-energy cosmic rays which by interaction with synchrotron photons produce ultra-high-energy neutrinos and gamma-rays. The typical "high-energy particle provider" hot-spot is thus an extended hot-spot with a strong enough magnetic field and a synchrotron cut-off lying at least in the optical band.

The neutrinos production criterion can then be applied to any hot spot displaying observational features in agreement with parallel shock acceleration. Such objects should display electrons distribution close to power-laws with indices close to $-\alpha_e - 2 = -3r/(r-1)$. Since it is believed that extragalactic jets are super fast-magnetosonic, the structure of the terminal shock is likely to be similar to a strong shock ($r = 4$) which naturally leads to $\alpha_e = -2$ and a synchrotron power-law index of $\alpha_s = (\alpha_e - 1)/2 = 0.5$.

In Meisenheimer et al (1997), authors present a sample of hot-spots which agrees with previous conditions. For these hot-spots, we can then apply our previous statements as shown in Tab. (1). The six hot-spots listed here show different characteristics which lead to various results. A trend correlating the synchrotron cut-off frequency and the maximum proton energy $E_{CR,max}$ does appear. The hot-spot size has also a large limiting impact on $E_{CR,max}$ as it is the case for the objects like 3C111 E and 3C20 W with large escape losses. We found only one object over six with enough accelerating capabilities to produce cosmic rays at energies of the order of $10^{20}$ eV. Both previous trends have to be confirmed with the help of a larger sampling, a work postponed to the future. 3C273 A is the best candidate for high-energy production since the maximum cosmic ray energy attainable is $7.9 \times 10^{19} eV$. Such a high-energy is well above the $p\gamma$ threshold energy associated with synchrotron emission and should thus lead to a high-energy neutrino emission. Looking back at Eq. (26) and (27) we make the conjecture that *the best hot-spot candidates for ultra-high-energy particle (cosmic rays, gamma-rays and neutrinos) production are extended ($L_{HS} \geq 1kpc$), not too weakly magnetized ($B > 0.1mG$) terminal shocks displaying synchrotron emission cut-off lying at least in the optical band.*
The FRII hot-spots are probably not the sources of the highest energy cosmic-rays detected on Earth both because there are no such strong sources within 50 Mpc (Elbert & Sommers, 1995) and because most of the hot-spots are not efficient cosmic-ray accelerators.

Forthcoming observations from neutrinos telescopes such as ANTARES should provide properties of the neutrino populations emitted from astrophysical objects such as hot-spots. The observed spectra should display boundaries that will test the acceleration mechanism of cosmic rays. Indeed, if cosmic rays reach energies where photo-meson production can occur in relation to synchrotron photons, the resulting neutrino spectrum will range from a minimum energy $\epsilon_{\nu,min}$ to a maximum one $\epsilon_{\nu,max}$. The energy of an emitted neutrino from $p - \gamma$ interaction is at the resonance approximately $\epsilon_\nu = 0.05 E_p$, or using Eq. (26) and (27)

<u>*Kolmogorov*</u>

$$\epsilon_{\nu,max} = 2.5 \times 10^{11} eV \left(\frac{h\nu_c}{1eV}\right)^2 \left(\frac{\zeta^{3/2}}{\beta_{jet}^3}\right) \left(\frac{B}{0.1mG}\right)^4 \left(\frac{L_{HS}}{1kpc}\right)^3 \quad (30)$$

<u>*Kraichnan*</u>

$$\epsilon_{\nu,max} = 1.4 \times 10^{11} eV \left(\frac{h\nu_c}{1eV}\right)^{3/2} \left(\frac{\zeta}{\beta_{jet}^2}\right) \left(\frac{B}{0.1mG}\right)^{5/2} \left(\frac{L_{HS}}{1kpc}\right)^2 . \qquad (31)$$

The minimum neutrinos energy is directly given by the synchrotron cut-off through Eq. (10) and is independent of the type of turbulence

$$\epsilon_{\nu,min} = 3 \times 10^{15} eV \left(\frac{h\nu_c}{1eV}\right)^{-1} . \qquad (32)$$

For instance in 3C273 A, the above mentioned neutrinos spectrum limits are such that $\epsilon_{\nu,min} \simeq 1.5 \times 10^{15} eV$ and $\epsilon_{\nu,max} = 4 \times 10^{18} eV$ (Kolmogorov) or $\epsilon_{\nu,max} = 7.5 \times 10^{15} eV$ (Kraichnan).

*3.3 Perpendicular shocks*

The magnetic structure of an extragalactic jet has two main components, namely the axial one and a toroïdal one. When the toroïdal component is much larger than the other, the terminal shock occurring in the hot-spot has a structure where the magnetic field is perpendicular to the shock normal. In that case, the magnetic field is compressed by a factor equal to the shock compression ratio r. Jokipii (1987) has first studied the effect of the magnetic field orientation on the diffusive shock acceleration efficiency. One of the main result was that the perpendicular shock configuration is always more efficient to accelerate particles than parallel shock. This estimation has been done in the framework of the quasi-linear theory which turns out to be accurate for parallel diffusion but more doubtful for perpendicular diffusion. We will thus extend the Jokipii (1987) study to all known diffusion regimes. The magnetic field used in this section is the averaged magnetic field seen by a particle crossing back and forth the shock. This magnetic field is related to the downstream magnetic field by

$$\bar{B} = B_d \sqrt{\alpha/r^2 + (1-\alpha)} , \qquad (33)$$

the factor $\alpha = (1 + r^\beta)^{-1}$ is the fraction of time the particle stays in the upstream medium. In order to get this expression we assumed that both the turbulence maximum scale and the turbulence level up and downstream are identical.
The estimates presented in the previous subsection are no longer valid since $D_\parallel$ is replaced by $D_\perp$. Following the same procedure as in the previous 2, we obtain the expression of the electron cut-off using Eq. (19)

$$\epsilon_{c,\perp} = \eta_T^{\frac{-2.3}{3-\beta}} \times \epsilon_{c,\parallel} \qquad (34)$$

where the subscripts $\perp$ and $\parallel$ stand for perpendicular and parallel shocks respectively. The corresponding synchrotron cut-off is then

$$h\nu_{c,\perp} = \eta_T^{\frac{-4.6}{3-\beta}} \times h\nu_{c,\parallel} . \tag{35}$$

The efficiency of the acceleration is larger in this kind of configuration since $\eta_T \leq 1$. From an observational point of view, all observed hot-spots present synchrotron cut-off frequencies smaller than $10^{16} Hz$. The observational constraint in relation with respect to synchrotron emission cut-off provides an important test for perpendicular shocks. Indeed rewriting Eq. (35) in terms of turbulence level gives

*Kolmogorov*

$$\eta_T = \left\{ \frac{\beta_{jet}^2}{\zeta} \left( \frac{4.06 \times 10^{18} Hz}{\nu_c} \right)^{2/3} \left( \frac{B}{0.1 mG} \right)^{-1} \right\}^{1/1.3} \left( \frac{\lambda_{max}}{1 kpc} \right)^{-2/3.9} \tag{36}$$

*Kraichnan*

$$\eta_T = \left\{ \frac{\beta_{jet}^2}{\zeta} \left( \frac{5.8 \times 10^{19} Hz}{\nu_c} \right)^{3/4} \left( \frac{B}{0.1 mG} \right)^{-3/4} \right\}^{1/1.3} \left( \frac{\lambda_{max}}{1 kpc} \right)^{-1/2.6} \tag{37}$$

We display in Tab. (2) the corresponding value of the turbulence level for all hot-spots presented in Tab. I. For these estimates we have set $\lambda_{max}$ equal to the size of the hot-spot which minimizes $\eta_T$. None of the hot-spots listed here, except 3C 273 A, fulfills $\eta_T \leq 1$ using Eq. (36) and (37) which proves that in these structures, the acceleration taking place in such terminal shocks cannot be consistent with a perpendicular shock structure. For the particular case of 3C 273 A, the synchrotron cut-off frequency lies in the optical range, contrary to all other hot-spots listed here betraying the presence of an efficient particle acceleration as for instance in perpendicular shock. We can then assume that such an emitter (weak radio power but significant optical emission) is likely to display terminal shock structure with dominant toroidal magnetic fields.

*3.4 Bohm regime*

Using the same procedure as in the two previous cases, we calculate the electron spectrum cut-off with $D = D_B = R_L c/3$. The result is then

$$\epsilon_c = 7.27 \times 10^{15} eV \frac{\beta_{jet}}{\zeta^{1/2}} \left( \frac{B}{0.1 mG} \right)^{-1/2} \tag{38}$$

| Hot-spot | $\eta_{T,min}^{\beta=5/3}$ | $\eta_{T,min}^{\beta=3/2}$ |
|---|---|---|
| 3C405 A | 2.1 | 32.7 |
| 3C405 D | 3.2 | 49.9 |
| 3C111 East | 11.1 | 99.3 |
| 3C20 W | 4.6 | 48.4 |
| 3C273 A | 0.33 | 4.1 |
| 3C123 E | 19.8 | 358.2 |

Table 2
Minimum values of the turbulence level for the cases where hot-spots have a perpendicular shock structure. According to its definition, this parameter is such that $\eta_T \leq 1$. Perpendicular shock acceleration can then only occur in 3C273 A for the case of a Kolmogorov turbulence.

which corresponds to a synchrotron cut-off frequency

$$\nu_c = 2.32 \times 10^{22} Hz \frac{\beta_{jet}^2}{\zeta} \ . \tag{39}$$

The restriction $\nu_c < 10^{16} Hz$ seen in all hot-spots, rules out the Bohm diffusion regime since it would imply that $\beta_{jet} < 7 \times 10^{-4} \zeta^{1/2}$. According to values of terminal jet velocities of FR-II jets, this latter relation is completely unrealistic for FR-II terminal shocks. The Bohm regime can be expected at shocks but it is likely that it is produced by a different type of turbulence (for instance, generated by streaming instability of relativistic particles).

In this regime, protons are confined in the source for a time-scale $t_{esc} \simeq L_{HS}^2/4D_B \sim 4\ 10^4\ B_{0.1}\ (E_{cr}/10^{19}\ \text{eV})$. As discussed by Aharonian (2002), for this time-scale to be shorter than the synchrotron loss time-scale, the product $B_{0.1}^3\ (L_{HS}/1\ \text{kpc})^2$ has to be $\gg 1$. This condition is mandatory in order to explain the diffuse X-ray emission seen in recent Chandra observations of extragalactic jets and hot-spots, which is produced by synchrotron radiation from relativistic protons. The assumption of a Bohm coefficient leads to a high value ($\geq 1$ mGauss) of the magnetic field on the whole source volume. The above result questions the validity of this assumption, at least for the hot-spots of FRII sources. The other turbulence scalings lead to shorter escape time-scales and reinforce this conclusion.

*3.5 Other MHD scalings*

The hot-spot is a high beta, high temperature medium, slightly super-Alfvénic. No work on chaotic transport in these regimes has been performed (difficult to conclude for quasi-perpendicular shocks), but the Alfvén and magnetosonic waves are not efficient in the diffusion process of high-energy particles compared to the isotropic cases. This leads to larger acceleration time-scales and lower maximum cosmic ray energies.

In order to illustrate these arguments, we have derived, following Chandran (2000), the parallel diffusion coefficient produced by an anisotropic *incompressible* MHD turbulence with a Goldreich-Sridhar power spectrum (see Goldreich & Sridhar (1995)). The high-energy electron Larmor radii in a few hundred $\mu$ Gauss magnetic field is $\simeq 10^{-5}$ parsec, much lower than the expected turbulence maximum scale $\lambda_{max}$. At low rigidities, the quasi-linear theory can be applied and particles are mostly accelerated through the transit-time damping (TTD) effect (magnetic mirroring effect). The parallel diffusion coefficient from (Chandran, 2000) is

$$D_\parallel \simeq (\frac{5}{2} - \frac{3\pi}{4}) \lambda_{max} \, c \, \frac{\beta_a}{\ln(1/\rho)} \, , \tag{40}$$

where, $\beta_a = V_a/c$ and $\rho = R_L/\lambda_{max}$.

This value can be injected into the acceleration time-scale (Eq.(17)) and compared to the synchrotron loss time leads to an estimate of the maximum turbulence length $\lambda_{max}$ in terms of observable quantities ($u_{jet}, B, E_c$)

$$\frac{\lambda_{max}}{1 \; kpc} = K_1 \, [K_2 + \ln(\frac{\lambda_{max}}{1 \; kpc})] \, , \tag{41}$$

with $K_1 = 10^3 \, \xi \, \beta_{jet}^{-2} \, B_{0.1}^3 \, E_c \, n_p^{-1/2}$, $K_2 = \ln(10^8 \, E_c \, B_{0.1})$. The magnetic field is expressed in 0.1 mG units, the electron cut-off energy $E_c$ is expressed in TeV and the proton thermal density $n_p$ in cm$^{-3}$ units. Eq.(41) leads to small turbulence maximum scales $\lambda_{max} \simeq 0.6 pc$ for standard hot-spot parameters. We verify a posteriori that the condition $\rho \ll 1$ does apply for the Eq.(40) to be valid. This value is naturally explained by the inefficiency of the scattering in an anisotropic turbulence. Such a low $\lambda_{max}$ does not allow high proton maximum energies. If we use this estimate and calculate the cosmic ray confinement energy by $R_L(E_{pc}) = \lambda_{max}$, we obtain $E_{pc} \sim 6 \; 10^{16} \, B_{0.1}$eV.

Yan & Lazarian (2002) discussed the cosmic ray transport in *compressible* anisotropic MHD turbulence and concluded that an overestimate of scattering frequency in incompressible models was made. The above conclusions are then likely to be optimistic for hot-spots and we may expect cosmic rays with energies lower than $10^{17}$ eV if such a MHD turbulence applies.

Real hot-spots may however harbor different turbulent regions where different kinds of scalings do apply. Quite possibly, the turbulence at the terminal shock is different from the turbulence at larger scales. This configuration would deserve specific investigations that are beyond the scope of the present paper and are thus postponed to the future.

# 4 Multi-scale numerical calculations and synthetic spectra

In this section we intend to illustrate the previous estimates about particle acceleration by presenting MHD-Kinetic numerical calculations of electrons and cosmic ray spectra produced in extragalactic jet terminal shocks, displaying different properties. We provide some support to in-situ particle acceleration and select the hot-spots by their ability to be described by diffusive shock acceleration. We then select for our computation two representative examples of hot-spots, one being a strong radio emitter without any optical emission (3C405 A) and the other one being a weak radio emitter with significant optical emission (3C273 A). We shall first constrain the turbulence properties in the Kolmogorov case by obtaining a realistic electron spectrum in agreement with observations and then we shall do the same simulations for cosmic rays acceleration and $p - \gamma$ and $p - p$ secondary particle production. We finally derive the contribution of the hot-spot of 3C273A class to the gamma-ray and high-energy neutrinos extragalactic backgrounds.

## 4.1 In-situ particle acceleration and transport

The large distance between the hot-spots and the galactic nucleus makes it unlikely that there is direct injection of relativistic particles from the nucleus, the radiative loss time-scales being too short compared to advection. Brunetti et al (2003) have detected a sub-sample of hot-spots in optical wave-bands with the VLT, this fraction represents a large part (up to 70 %) of the total sample of 10 objects. The shape of the synchrotron spectrum emitted by this sub-sample hot-spot is consistent with particle acceleration within a low magnetic field. In high loss hot-spots spectral breaks in the flux $\Delta\alpha \simeq 0.5$ are observed at $\sim 10$ GHz, consistent with diffusive shock acceleration at the terminal shock and strong downstream radiative losses in high magnetic fields (see Carilli et al (1999) and references therein for the well studied case of Cygnus A).
However, diffusive shock acceleration at the terminal shocks is difficult to conciliate with *diffuse* optical emission downstream as observed in Pictor A-W (Perley et al, 1997). Other hot spots like 3C33-S do show hard spectral indices not easily explained within the diffusive shock acceleration framework (Meisenheimer et al, 1997). These more complex configurations deserve special investigation postponed to future works. We therefore disregarded objects like Pictor A-W or 3C33-S from our sample (see Tab. I in Section 3).

## 4.2 SDE validity and rescaling method

In order to apply the SDE formalism to a problem, it has been shown that the physical system has to fulfill a relation involving the advection length $\Delta X_{adv} =$ $V_{adv}\Delta t$, the diffusion length $\Delta X_{diff} = \sqrt{2D\Delta t}$ and the thickness of the shock $\Delta X_{sho}$ (Krülls & Archterberg, 1994; Casse & Marcowith, 2003). This relationship implies that in our computation

$$\Delta X_{adv} \ll \Delta X_{sho} < \Delta X_{diff} . \tag{42}$$

For the case of MHD simulations, the shock thickness is determined by the size of a grid cell or in adaptative mesh refinement (AMR) simulations the size of the smallest grid cell. This double inequality can be reformulated in terms of conditions pertaining to the diffusion coefficient and the SDE time step, namely

$$D > D_{min} = \xi_{adv}\xi_{diff}^2 \frac{|V_{adv}|\Delta X_{sho}}{2} \; ; \; \Delta t \leq \frac{\Delta X_{sho}}{\xi_{adv}|V_{adv}|} \tag{43}$$

where $\xi_{adv,diff} > 1$ since $\Delta X_{diff} = \xi_{diff}\Delta X_{sho}$ and $\Delta X_{adv} = \Delta X_{sho}/\xi_{adv}$. The spatial resolution of the MHD numerical calculation then plays a crucial role in the validity of the use of SDE. In the present work, we deal with AMR computation having an initial grid size of $100 \times 40$ and a refinement acting on five sub-levels leading to an actual grid resolution of $3200 \times 1280$. The actual axial spatial resolution is then of the order of $\Delta X_{sho} \sim 10^{-3} kpc$. Since diffusion coefficients generally depend on particle energy, the above relation can be reformulated in terms of a minimal particle energy. For the case of Kolmogorov turbulence occurring in the vicinity of a parallel shock, the particle energy threshold for SDE validity is

$$\frac{\epsilon}{1\,GeV} > 1.23 \times 10^4 \left(\beta_{jet}\eta_T\right)^3 \left(\frac{\Delta X_{sho}}{10^{-3}kpc}\right)^3 \left(\frac{B}{0.1mG}\right)\left(\frac{\lambda_{max}}{1kpc}\right)^{-2} . \tag{44}$$

In our calculations, we always set the maximal turbulence wavelength $\lambda_{max}$ equal to the size of the hot-spot. This leads us to infer directly the turbulence level $\eta_T$ in these objects by considering both the synchrotron cut-off frequency and observational properties. For 3C405 A, a parallel shock sustaining an MHD Kolmogorov turbulence leads to $\eta_T = 0.31$. Injecting this value into the SDE particle energy threshold gives $\epsilon_{min} = 5 GeV$. If cosmic ray dynamics agrees with this constraint, kinetic electron computation will be difficult to achieve since the cut-off energy is of the order of $70 GeV$. One way to overcome this problem is to "re-scale" the diffusion coefficients, namely to artificially increase their value by a certain factor. Doing this, we artificially increase the Fermi acceleration time, so that in order to obtain the accurate spectrum we thus need to also artificially increase the synchrotron time-scale by the same factor. This method is valid for computing the energy electron spectrum at the shock because diffusion and synchrotron emission are not related. For the case of cosmic rays, this trick would not be accurate since both Fermi acceleration and particle leakage both depend on diffusion coefficients.

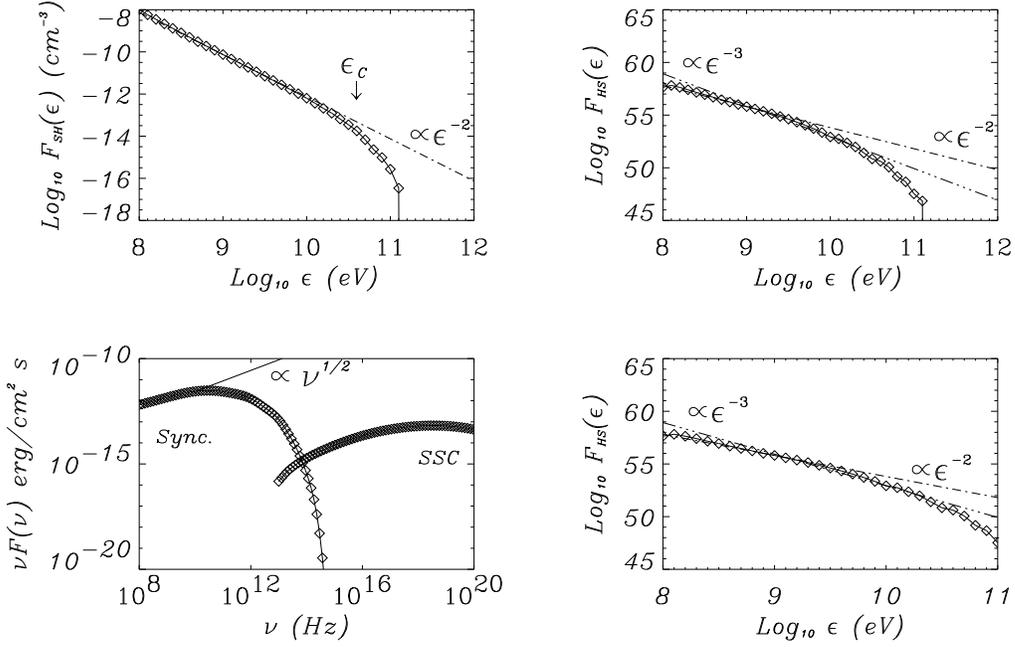

Fig. 3. Energy spectra of electrons accelerated at the terminal shock of a jet simulated by a MHD simulation displaying all observational properties of one of Cygnus A hot-spots (3C405 A). A MHD Kolmogorov turbulence is assumed to occur. In the upper left panel, the electron energy spectrum at the shock displays two characteristic behaviors, the first one being a power law regime of index $-2$ corresponding to a strong shock acceleration and the second one, beyond the energy $\epsilon_c$ calculated with Eq. (20), being a rapid decrease of the number of particles corresponding to a synchrotron cut-off. In the right panels, we have displayed the electron distribution function integrated over the whole hot-spot. Its behavior is different from the previous one since during their propagation within the hot-spot, electrons are prone to synchrotron losses which lead to a shift in the power-law index from $-2$ to $-3$. In the lower left panel, we have displayed the synchrotron and the synchro-Compton spectra emitted by the hot-spot by the electrons. For the synchrotron spectrum we find a power law regime as well as a frequency cut-off in agreement with observations. For the synchro-Compton spectrum the flux at few keV is found to be consistent with the flux reported by Chandra (Wilson et al, 2000)

*4.3 Cygnus A (3C405A)*

The choice of this hot-spot has been made according to its observational properties. It is indeed a typical illustration of a radio-loud jet terminal shock class whose spectrum cut-off frequency lies in the infra-red band. As seen in the previous section, this kind of shock is probably a poor particle accelerator and is very unlikely to be a source of ultra-high-energy cosmic rays. Scanning various types of known diffusion regimes (Kolmogorov, Kraichnan, Bohm), we have inferred from the observational source features that the terminal shock is likely to be a magnetically parallel strong shock. In order to obtain an MHD simulation close to the topology of this hot-spot, we adopted the initial MHD conditions described in Sect. (2.3) and set the toroidal magnetic field to zero so that the shock structure is close to a parallel shock. We also adopted the quantities provided by Meisenheimer et al (1997), namely $U_{jet} = 0.24c$ and $B = 0.4mG$ as velocity and magnetic field values. The only unknown physical quantity at this stage is the radius of the downstream extragalactic jet. We have set its value to $100pc$, a typical value for extragalactic radii.

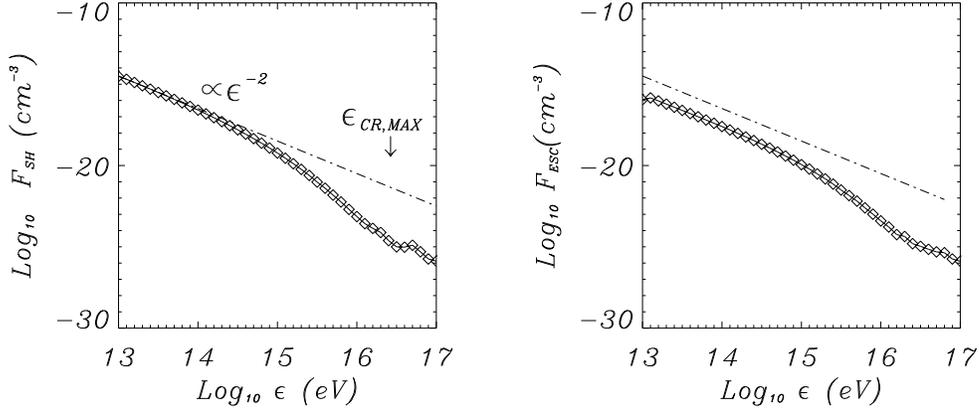

Fig. 4. Energy spectra of cosmic rays accelerated using the same conditions as in Fig. 3. In the left panel, we show the cosmic ray spectrum measured at the shock front and in the right panel, the spectrum of escaped particles. Once again we obtain a two feature shock spectrum, namely at low energies a strong shock-like power law and at energies close to the cut-off energy a rapidly decreasing curve. The cut-off energy $\epsilon_{CR,max}$ is calculated from Eq. (26) where the dominant energy loss mechanism is the particle leakage. Near this cut-off energy the spectra measured at the shock front and at the hot-spot boundaries are the same, which show that all particles accelerated up to $\epsilon_{CR,max}$ are rapidly escaping from the hot-spot.

The temporal evolution of the hot-spot is rather slow compared to the particle acceleration time-scale. Indeed the evolution time-scale of the hot-spot can be considered as $t_{HS} \sim L_{HS}/U_{HS}$. In the MHD simulation described in the previous section (Fig. 1), the terminal shock reaches a ballistic motion whose propagation velocity is much smaller than the jet velocity (see Fig. 2 $U_{HS} = 0.17 U_{jet}$). For the case of 3C405 A, the hot-spot evolution time-scale is then

$$t_{HS} \sim 2.5 \times 10^{12} s \left(\frac{L_{HS}}{1kpc}\right) \left(U_{jet}/0.24c\right)^{-1} . \tag{45}$$

Comparing it to the Fermi acceleration time for a parallel shock, we obtain that $t_{FI} < t_{HS}$, provided that

$$\epsilon < 10^{19} \left(\frac{B}{0.1mG}\right) \left(\frac{\lambda_{max}}{1kpc}\right)^{-2} \left(\frac{L_{HS}}{1kpc}\right)^3 \left\{\frac{62.5 \beta_{jet} \eta_T}{\zeta}\right\}^3 eV . \tag{46}$$

For 3C405 A, this upper limit corresponds to $\epsilon < 8 \times 10^{17} eV$. According to previous estimates of cosmic ray acceleration, this condition is always fulfilled so we can safely use a single snapshot of the MHD simulation to compute the kinetic calculation.

The selected MHD snapshot is the one represented in Fig. 1 at $T = 140$ because it belongs to the ballistic motion range and fulfils the above statement. The terminal shock displays all characteristics of a strong shock where magnetic field lines are perpendicular to the shock front. The result of the transport of a mono-energetic electron population injected near the shock at an initial energy of $100 MeV$ is shown on Fig. (3). On the upper left panel, we have displayed the distribution function of these electrons measured at the shock front. The spectrum exhibits an expected behavior since for $\epsilon < \epsilon_c = 70 GeV$ (i.e. energies where synchrotron losses are weak, $\epsilon_c$ being calculated from equation (20)), we obtain a power law whose index is equal to $-2$. As shown in equation (5), this is exactly the shape of the distribution function expected from a strong shock ($r = 4$) where all energetic losses can be neglected. On the other hand, when $\epsilon > \epsilon_c$, the spectrum rapidly decreases (as the spectrum slope), showing a synchrotron cut-off feature. In the right panels, we have displayed the electron energy distribution integrated over the whole hot-spot volume. The obtained spectrum takes into account the synchrotron cooling occurring during the particle propagation leading to a spectral break $\Delta s = 1$ above $E_b$ in the particle distribution function.

In the last panel, we have computed the corresponding synchrotron spectrum emitted from the whole hot-spot by the electrons. The synchrotron emission is mainly achieved in the radio and infra-red domain ($\nu < 10^{13} Hz$) with an intensity behaving as a power law of spectral index 0.5, followed by a plateau (corresponding the synchrotron frequencies emitted by electrons of energy $E_b$) up to a cut-off frequency of $\nu_c = 9 \times 10^{12} Hz$, in agreement with 3C405 A observations (Meisenheimer et al, 1997). We interpret the X-ray emission reported by Chandra as synchro-Compton radiation, alternative models as proton initiated cascade (Mannheim et al, 1991) or Inverse Compton radiation on cosmic-microwave background photons being non relevant here: the maximum proton energies are under the pion production threshold and the magnetic field energy density dominates over the CMB field energy density. We leave for a future work the detailed spectral and spatial analysis of this high energy component.

The electron normalization $N_0$ is obtained using the measured flux at 5 Ghz reported and the energy equipartition magnetic field (see tables 5 and 6 in Meisenheimer et al (1997)). The density is then calculated using the synchrotron emissivity $\epsilon_\nu$ (see Pacholczyk (1970)). Once the electron density is known, the relativistic electron energy density is given by

$$U_{re} \simeq N_0 \, (ln(E_b/E_{min}) + \frac{1}{E_b}) \, m_e \, c^2, \tag{47}$$

where $E_{min}$, $E_b$ are the minimum and break energy (in eV) of the electron distribution respectively. The magnetic field energy density is

$$U_{Bme} = \frac{3}{4} U_{re}(1 + k_p),  \qquad (48)$$

where $k_p$ is the ratio of relativistic proton to relativistic electron energy density. This last relationship allows us to estimate the proton energy density, in order to normalize our proton spectrum (see next section).

For 3C405A, the relativistic electron density is found to be of the order of $10^{-4} cm^{-3}$ which leads to an electron density at 100 MeV of $10^{-8} cm^{-3}$ (derived from the power law electron spectrum $N(\gamma_e) \propto \gamma_e^{-2}$). The electron energy density is then, according observations of Meisenheimer et al (1997), $U_{re} \sim 6 \times 10^{-10} erg/cm^3$ ($E_b/E_{min} \sim 1800$). Considering the minimum value of the magnetic field of 3C405 A ($B \sim 0.35 mG$), we easily obtain the cosmic ray energy density, namely $U_{CR} \sim 8 \times 10^{-9} erg/cm^3$.

Confident in our choice of the hot-spot parameters, we have then computed the kinetic transport of cosmic rays in the same way as before except for the injection energy which is shifted from $100 MeV$ to $10^{13} eV$. This choice is imposed by numerical limitations, since a power law spectrum with a spectral index equal to $-2$ requires us to consider more than a million particles in order to achieve the computation over three decades in energy. We must also specify that we have imposed that if a particle is at a location distant by more than $L_{HS}$ from the shock, we consider this particle to have escaped. This condition (also imposed for the previous electron computation) plays a crucial role for cosmic rays since it is expected that particle leakage is the dominant energy loss mechanism. This is proved by the spectra displayed in Fig. 4 where on the left panel we have represented the cosmic ray spectrum measured at the shock front and in the right the cosmic ray measured at the hot-spot boundaries, namely at the location where a cosmic ray has escaped from the source. These two spectra differ at low energies ($\epsilon \ll \epsilon_{CR,max} = 4.1 \times 10^{16} eV$, c.f. Eq. (26)). Indeed, while at the shock front the cosmic ray spectrum exhibits the expected power law shape, the escaped cosmic ray spectrum is curved with a lower normalization, betraying a very good particle confinement at low energy. On the other hand when $\epsilon \sim \epsilon_{CR,max}$, the cosmic ray spectrum at the shock front significantly differs from the power law and becomes comparable in shape and normalization to the escaped cosmic ray spectrum. This is direct evidence that particles can no longer be confined in the hot-spot when $\epsilon \geq \epsilon_{CR,max}$ and that this hot-spot cannot produce cosmic rays beyond this maximal energy. The normalization of the cosmic ray spectrum has been obtained by using the previous estimate for the cosmic ray energy density $U_{CR}$ since

$$U_{CR} = N_{o,CR} ln\left(\frac{\epsilon_{CR,max}}{\epsilon_{min}}\right) m_p c^2 . \qquad (49)$$

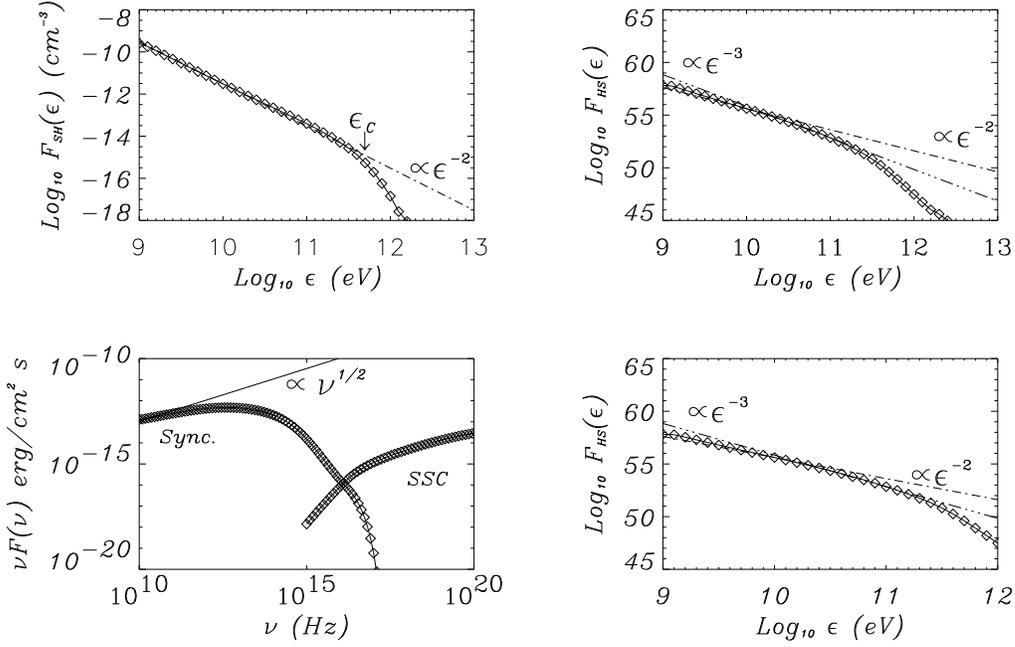

Fig. 5. Same figure as Fig. 3 but applied to the hot-spot 3C273 A. In this hot-spot a toroidally-dominated magnetic structure is likely to occur so that the electron spectrum (upper left panel) extends up to $\epsilon_c = 500 GeV$, leading to synchrotron emission (lower left panel) extending to the optical band where $\nu_c \sim 3 \times 10^{14} Hz$. The electron distribution is obtained by first inferring the turbulence level from the observational synchrotron cut-off frequency and then reproducing the synchrotron emission features by means of kinetic computation. As for Cygnus A, we also provide for the synchro-Compton emission falling in the X-rays to be expected by the hot-spot.

We obtain then an average cosmic ray density of $N_{o,CR} \sim 3 \times 10^{-7} cm^{-3}$.

To conclude on Cygnus A, we have shown that our numerical simulations can reproduce the multi-wavelength observations well: the break and cut-off frequencies of the synchrotron spectrum, the spectral indices under and above $E_b$, the X-ray spectrum produced by SSC with a magnetic field close to equipartition among relativistic electrons and protons. The hot-spot magnetic configuration is likely dominated by its poloïdal magnetic and the particle transport controled by a Kolmogorov-type turbulence.

### 4.4  3C273 A

The second hot-spot we have selected belongs to a different class of hot-spot. This object exhibits interesting physical properties for particle acceleration. Indeed, as seen in the previous section, its cut-off synchrotron frequency lies in the optical band while its size and magnetic field amplitude remain of the same order as in Cygnus A. As seen in the analytical estimates of electron cut-off energy in both perpendicular and parallel shock regimes, this hot-spot is likely to have a magnetic perpendicular structure where the toroidal magnetic field component is much larger than the poloidal one (see Tab. (II)). The resulting smaller spatial diffusion coefficient across the shock front naturally leads to a higher cut-off energy for electron distribution. The same reasoning can also be applied to cosmic ray acceleration, since the size of the hot-spot is of the same order as for Cygnus A. As seen in Tab. (I), the cosmic ray maximal energy attainable within 3C273 A is expected to be beyond $10^{19} eV$ in the Kolomogorov turbulence regime, so that one can expect pion photo-production to occur within this source.

We have adopted the same approach as in the previous paragraph, except for the chosen MHD simulation. We have performed an MHD simulation where initial conditions include observational values of jet velocity ($U_{jet} = 0.27c$) and magnetic field ($B = 0.35 mG$) as well as a magnetic structure where the toroidal component dominates the poloidal one ($B_1 = 100 B_o$). The resulting MHD structure is very similar to the previous one in terms of density, velocity field and thermal pressure; the only real difference being the relative amplitude between magnetic components. In order to check if we can use a single snapshot of the structure or a time-coupled MHD-kinetic simulation, we have calculated the energy at which the first-order Fermi acceleration time would exceed the typical evolution time-scale of the hot-spot Eq. (46) and we find that in the perpendicular diffusion regime, this maximal energy is

$$\epsilon < \frac{4.3 \times 10^{19} eV}{\eta_T^{3.9}} \ . \tag{50}$$

Assuming the maximal turbulence wavelength to be equal to the size of the hot-spot, we can infer the value of the turbulence level $\eta_T$ in order to make the theoretical synchrotron frequency matching the observational one. This values is equal to $\eta_T = 0.4$ for the case of 3C273 A so that the above "snapshot" criterion is always fulfilled, since $\epsilon_{CR,MAX} = 7.9 \times 10^{19} eV < 1.5 \times 10^{21} eV$.

Using one snapshot of the MHD structure and applying the kinetic scheme for the acceleration of electrons, we get the electron spectrum in the hot-spot shown in the top panel of Fig. 5, where we have injected a million particles at an initial energy $\epsilon = 1 GeV$ for a total simulation time of 2000 (in units of $r_{jet}/v_{jet} \simeq 10^3$ yrs). The shape of the spectrum is the same as in the previous section apart from the location of the energy cut-off which lies near $500 GeV$, corresponding to the expected synchrotron cut-off frequency near $4 \times 10^{14} Hz$ (see lower panel of Fig. 5). The radiative (synchrotron) losses induce a break in the particle distribution which steepens to $E^{-3}$ above $7-8 \ 10^3 m_e \ c^2$. The observed synchrotron break corresponds to a few $10^3 m_e \ c^2$ and would have required a much longer simulation, but as the minimum energy of the interacting photons in the hadronic process is $\sim 10^{11} Hz$, we stopped our simulations at $t = 2000$.

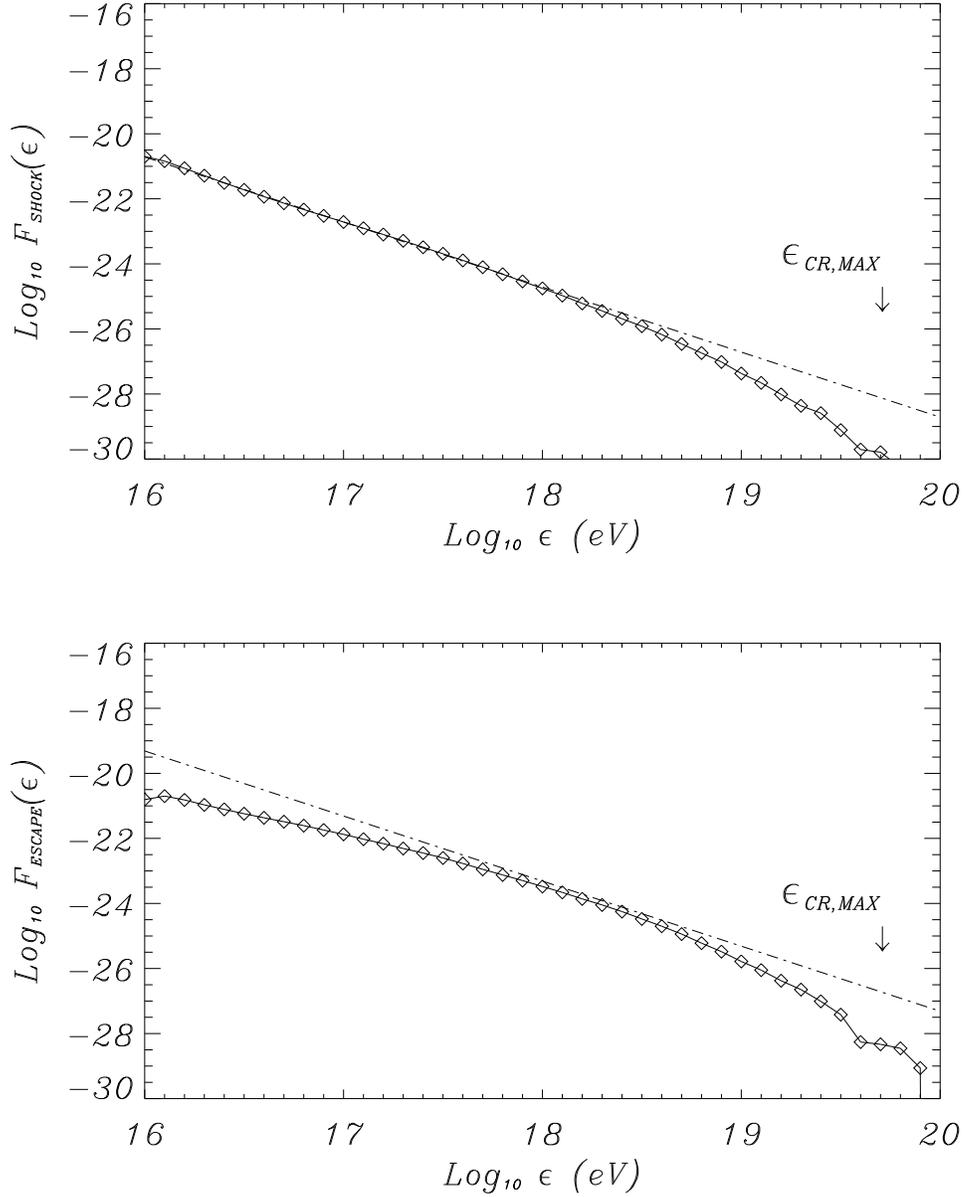

Fig. 6. Proton spectrum for the 3C273A hot-spot. Upper panel: The proton distribution (in $cm^{-3}$ units) at the terminal shock. The energy spectrum scales as $E^{-2}$ up to $E_{CR} \simeq 10^{19}$ eV where escape losses are dominant. Lower panel: The proton distribution escaping the hot-spot. At energies lower than a few $10^{17}$ eV the diffusive length of the particles is smaller than the hot-spot size and the spectrum hardens.

The distribution normalization is derived in a similar way as that for Cygnus A. 3C273A is however a less powerful radio emitter with $S_\nu(5Ghz) \simeq 2.1 Jy$. A energy equipartition magnetic field $B_{me} \simeq 0.35 mG$ gives a mean electron density of $N_0 \simeq 2.5\ 10^{-4} cm^{-3}$. The particle energy corresponding to the spectral break is $E_b \simeq 1.9\ 10^3\ m_e\ c^2$. The minimum energy of the distribution is an unknown, but if we take it in the range $1-10$ $m_e c^2$ the ratio $ln(E_b/E_{min})$ in Eq. (47) does not vary by more than a factor 2. With this error in mind, we finally get $U_e \simeq 7\ 10^{-10}$erg/cm$^3$. We also display the X-ray spectrum by SSC. The flux is about two orders of magnitude lower than the Cygnus one. This is due 1/ to a larger distance of the source 2/ to an intrinsic weakness of the flux 3/ to a peak of synchrotron spectrum shifted towards higher frequencies.

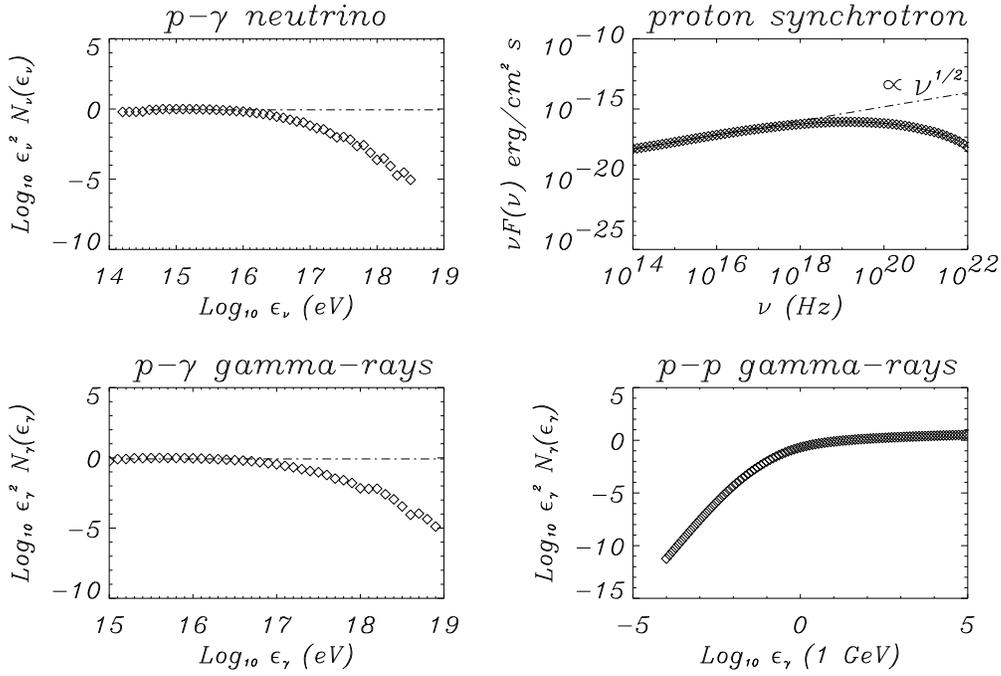

Fig. 7. Numerical astroparticle spectra produced in the hot-spot 3C273 A. The upper left panel represents the ultra-high-energy p-$\gamma$ neutrino spectra. The three other plots show the proton synchrotron (upper right), p-$\gamma$ gamma-rays (lower right) and p-p gamma-rays (lower left) ($N(\epsilon_{\nu/\gamma}) = F(\epsilon_{\nu/\gamma})$). Apart from the synchrotron flux, all other fluxes are normalized to their value on Earth (see text for these values). The neutrino and $\gamma$-ray yields come from the pions photo-production induced by cosmic rays interacting with synchrotron photons arising from the electron acceleration computed in Fig. 5. These two spectra exhibit a power-law behavior at low energies part coming from the combination of power-law cosmic ray and electron distribution interaction through the $\Delta$-resonance. (see text for more details).

Reproducing observational features such as the electron emission spectral slope and the cut-off frequency enables us to compute the cosmic ray transport and acceleration in the turbulence configuration within the limits of our numerical method. The main limit of our method is the test-particle hypothesis that may be a fair approximation if the cosmic-ray pressure tends to be a substantial fraction of the thermal pressure. However, if backreaction effects can lead to a different, for instance concavely shaped spectrum, for a given turbulence downstream (in the hot-spot) the maximum proton energy is always fixed by the geometry of the hot-spot and the magnetic field configuration. This maximum energy is expected to be of the same order as in the test particle case. However, a more realistic model would require to take into account such backreaction effects. Unfortunately this is beyond capabilities of modern computers (at least in 3D).

To achieve this computation, we conserve the same setting of spatial diffusion coefficients and hot-spot boundaries which account for particle leakage as soon as particles are located at a distance larger than $L_{HS}$ from the terminal shock. In Fig. 6 the upper panel represents the energetic cosmic ray spectrum where we have injected a million particles with initial energy $\epsilon = 10^{16} eV$. The resulting spectrum begins with a power-law with spectral index $-2$ in agreement with strong shock Fermi acceleration where no energetic losses are significant. Near the expected maximal energy $\epsilon_{CR,MAX} = 7 \times 10^{19} eV$, the spectrum slope is monotonicaly decreasing which can be explained as particle leakage becoming significant, such that the escape time is of the order of the first-order Fermi acceleration time. Reaching such high energies (up to $10^{20} eV$), cosmic rays produce secondary particles during their transport. Indeed, the electron synchrotron emission produces source photons whose energy range from radio up to the optical band ($h\nu_{max} \sim 1.7 eV$). This means that all cosmic rays whose energy is beyond $3 \times 10^{16} eV$ are interacting with these photons by the way of pion photo-production.

The cosmic ray spectrum normalization is given by Eq (48). The above energy densities $U_{Bme}$ and $U_{re}$ give, $k_p \simeq 8$ or a relativistic proton energy density $U_{pr} \simeq 5.6~10^{-9}$ erg/cm$^3$. For a $E^{-2}$ spectrum, the relativistic proton density is $N_p \sim 2~10^{-7}$ cm$^{-3}$, $\sim 10^{-5}$ times the mean thermal proton density in the hot-spot. The energy density of the *interacting* protons (with $\gamma_p \geq 7~10^7$) is $u_{pi} \simeq 2~10^{-9}$ erg/cm$^3$. The proton initiated cascade models (see Mannheim et al (1991)) predict that the X- and gamma-ray parts of the electromagnetic spectrum are produced by the radiation of secondary electrons. The predicted X-ray spectrum is harder compare to SSC and a nice test to discriminate among the two models would be an X-ray detection of the hot-spot (see the discussion concerning the hot-spot of 3C120 by Hardcastle et al (2001)). Here the proton luminosity is $L_p \simeq u_{pi}/t_{p\gamma} \simeq 2~10^{40}$ erg/s and following Mannheim et al (1991), we found a ratio of pion to SSC luminosity $L_\pi/L_{ssc} \simeq 10^{-6}$ for $u_{sync}/u_B \simeq 10^{-3}$. We do not then expect the PIC to dominate over the SSC process for 3C273A. Photohadronic processes do produce also neutrons through the channel $\gamma + p \rightarrow n + \pi^+$. Neutrons can escape and further decay on lengths $\ell_n \simeq \gamma_n~10^{-5}$ pc, that can be $\geq L_{HS}$ for $E_n = \gamma_n m_n c^2 \geq 10^{17} eV$. The neutron loss effect can be substantial on the proton spectrum near the maximum proton energies if the ratio $q~\tau_{esc}/\tau_{rad} \geq 1$. Here $\tau_{esc/rad}$ is the proton escape and dominant radiative (synchrotron or photohadronic) loss timescales respectively (see Biermann & Strittmatter (1987)). The parameter q ($\leq 1$) is the relative efficiency of the neutron channel written above. Unless the magnetic field and/or the photon field energy density being very high the fastest loss time for the protons is the spatial escape, hence $\tau_{esc} \leq \tau_{rad}$ and $R \leq 1$. The effect of neutrons is found to be negligible.

For completeness, we have computed the energy spectrum fluxes of both neutrinos and $\gamma$-rays coming from p-$\gamma$ interaction in the left panels of Fig. 7 accordingly to Eq. (12). These spectra have similar shapes, as expected from Eq. (12), and energies ranging between $10^{15}$ up to a few $10^{18} eV$ for neutrinos and from $10^{15} eV$ up to $10^{19} eV$ for $\gamma$-rays. Note that only the lower energy emission between $10^{15} eV$ and a few $10^{17} eV$ are really significant, the remaining upper energy neutrinos and $\gamma$-rays being a rapidly decreasing contribution. This is a feature induced by the cosmic ray spectrum which is dissociating from the regular power-law to a rapidly decreasing curve. Regarding the lower energy neutrino and $\gamma$-ray contribution, they agree with a power-law with a spectral index $-1$. This result can be derived using Eq. (12) evaluated at the $\Delta$ resonance given by the condition (10).

Close to the neutrino production threshold, at $\epsilon_\nu = 10^6$ GeV, for 3C273A, with $s_p = 2$, and a total photon density $N_{ph0} \simeq 10^{11} \text{ph/cm}^3$ and a neutrino flux on Earth of $\epsilon_\nu^2 \, dF_\nu/dt \simeq 10^{-15}$ GeV/cm$^2$ s sr. This flux is much lower than the best neutrino telescope sensitivity, about 5 $10^{-8}$ GeV/cm$^2$ s sr for ANTARES. However, the contribution of all 3C273A-like sources to the neutrino background could be susbtantial, but remains to be evaluated quantitavely.

In the lower left panel, we have displayed the synchrotron emission arising from the cosmic rays. Since the Larmor frequency of cosmic rays is much smaller than for electrons, this spectrum is 1/ shifted by a factor $m_{CR}/m_e$ in frequency (here plotted for protons) and 2/ the emissivity $F(\nu)$ is smaller by a factor $(m_e/m_{CR})^3$. This latter property makes this cosmic ray high-energy synchrotron emission almost impossible to detect considering moreover that the cosmic ray density is much smaller at ultra-high-energies than relativistic electrons at $GeV$ energies.

We finally calculate the p-p gamma-ray spectrum displayed. The target proton density is in our case of the order of $\simeq 10^{-2}$ cm$^{-3}$, most of the interaction occuring around the shock. We assumed a non relativistic proton injection energy $\leq 1 GeV$. The gamma-ray spectrum thus peaks at $\simeq 70 MeV$ and has a $E^{-2}$ spectrum above. With the aformentioned thermal and relativistic proton densities, the energy flux expected on Earth for 3C273A is $\epsilon_\gamma^2 \, F(\epsilon_\gamma) \simeq 4.7 \, 10^{-14}$ GeV/cm$^2$ s sr about three orders of magnitude under the GLAST sensitivity (5 sigma in 50 hours), and four orders of magnitude under the Tcherenkov telescopes sensitivities at 100 GeV (5 sigma in 50 hours). Considering, both the Inverse Compton and pp gamma-ray fluxes in the regime 1 GeV-1 TeV, one can conclude that the 3C273A class hot-spot is not predict to be a gamma-ray source for the present and the future generation of gamma-ray telescopes.

To conclude on 3C273A, as in the case of Cygnus A, our numerical simulations can reproduced the multi-wavelength observations well. We suspect the magnetic field configuration to be dominated by its toroïdal component, leading to lower acceleration timescales and higher particle energies. 3C273A is expected to produce high energy cosmic rays (up to $10^{20}$ eV) and high energy neutrinos and gamma-rays from the photo-pion production process. However, the level of astroparticle flux is low and not expected to be detected by the most sensitives future experiments.

## 5  Outlook

Following the pioneering work of Rachen & Biermann (1993), we have investigated, in some detail, the different cosmic ray transport regimes that may occur in FRII radio-galaxy hot-spots. We have first provided analytical calculations of the particle acceleration capability of hot-spots regarding cosmic rays and secondary particles such as $\gamma$-rays and ultra high-energy neutrinos in various magnetic configurations. One of our main conclusion is that the best hot-spot candidates for ultra-high-energy particle production are extended ($L_{HS} \geq 1 kpc$), rather strongly magnetized ($B > 0.1 mG$) terminal shocks displaying synchrotron emission cut-off lying in and above the optical band. Among our list of hot-spots, 3C273A appears to be one of the best candidates for high-energy astroparticle yield in hot-spots since it fulfills all aforementioned characteristics. However, 3C273 A is the only hot-spot in our list found to produce high energy cosmic rays.

As a second step, we have used a multi-scale approach based on the use of coupled MHD and kinetic numerical calculations to accurately compute relativistic electron and proton spectra. The kinetic scheme, using the stochastic differential equations (SDE) method, appears to be a simple and efficient way to solve complex kinetic problems taking into account the large-scale behavior of turbulent flows. Our approach enables us to accurately compute the spatial transport of both non-thermal electron and cosmic rays everywhere within the hot-spot and not just in the close vicinity of shocks. This differs from studies by Jones et al. (2002) where non-thermal electrons are simply advected by the flow between shock regions, thus favoring a time-dependent tracing of the flow structure.

The numerical calculations were done on the two different types of hot-spot discussed above, namely Cygnus A and 3C273 A. The numerical spectra that we have obtained confirm all analytical estimates done in this paper and enhance our prediction for the nature of the best hot-spot candidates for high-energy astroparticle yield. Moreover the study of the acceleration capacities of each type of hot-spot has led us to identify one main difference between these terminal shocks, namely the best particle accelerators are likely to arise from shocks whose dominant magnetic component is parallel to the front shock while the others are likely to have a dominant magnetic component perpendicular to the front shock. Lastly the high-energy p-$\gamma$ neutrino and p-p gamma-ray fluxes expected from hot-spot like 3C273A have been calculated but in our estimates these particle are not suitable for detection by the most sensitive observatory facilities available nowadays, partly because this type of source is far too distant from Earth (none of them are present within 50 Mpc around the Earth) and also because the proton luminosity is not expected to dominate over the electron one. Nevertheless, the estimated contribution done here may help to determine the contribution of this type of source to the cosmic diffuse neutrino and gamma-ray backgrounds expected to be detected in the forthcoming years.

*We warmly thank Rony Keppens for his helpful remarks and comments as well as for his hospitality during the stay of A.M. at FOM. F.C. is a fellow of the European Community's Human Potential Programme under contract HPRN-CT-2000-00153, "PLATON". We thank the referee for useful comments that led to the improvement of the manuscript. N.A. Webb is thanked for her careful reading of the manuscript.*